\documentclass[usenatbib]{mn2e}
\usepackage{epsfig}
\usepackage{amsmath}
\title[Low X-Ray Luminosity Clusters at z=0.25]{Galaxy Properties in Low X-Ray Luminosity Clusters at z=0.25}
\author[Balogh \etal]{Michael Balogh$^{1,2}$, R.\,G.\ Bower$^{1,4}$, Ian Smail$^{1}$, B.\,L.\ Ziegler$^{3,4}$, Roger L.\ Davies$^{1}$,
\newauthor A. Gaztelu$^{1}$, Alexander Fritz$^{3}$\\
$^{1}$ Department of Physics, University of Durham,South Road, Durham DH1 3LE, UK\\
$^{2}$ email: M.L.Balogh@Durham.ac.uk\\
$^{3}$ Universitaetssternwarte, Geismarlandstr. 11, 37083 Goettingen, Germany\\
$^{4}$Visiting astronomer of the German--Spanish Astronomical Center,
Calar Alto, operated by the Max--Planck--Institut f\"ur Astronomie,\\
Heidelberg, jointly with the Spanish National Commission for Astronomy.
}

\date{\today}
\def\etal{{ et al.\thinspace}}
\def\gtrsim{\mathrel{\raise0.35ex\hbox{$\scriptstyle >$}\kern-0.6em
\lower0.40ex\hbox{{$\scriptstyle \sim$}}}}
\def\lesssim{\mathrel{\raise0.35ex\hbox{$\scriptstyle <$}\kern-0.6em
\lower0.40ex\hbox{{$\scriptstyle \sim$}}}}

\def\h50{h_{50}}
\def\lowlx{Low--$L_X$}
\def\oiifull{[O{\sc ii}]$\lambda$3727}
\def\oii{[O{\sc ii}]}
\def\ha{H$\alpha$}
\def\hd{H$\delta$}
\def\ewoii{$W_\circ$([O{\sc ii}])}
\def\ewha{$W_\circ(H\alpha+$[N{\sc ii}])}
\def\ewhd{$W_\circ(H\delta)$}
\begin{document} 
\maketitle
\begin{abstract}
We present the first spectroscopic survey of intrinsically low X-ray luminosity clusters
at $z\gg0$, with {\it Hubble Space Telescope (HST)} WFPC2
imaging and spectroscopy from Calar Alto and WHT-LDSS2.  We study 
172 confirmed cluster members in a sample of ten clusters at $0.23<z<0.3$, with $L_X\lesssim 4\times 10^{43} h^{-2}$ ergs s$^{-1}$ [0.1-2.4 keV]
($\Omega_m=0.3,\Lambda=0.7$).  
The core of each cluster is imaged with WFPC2
in the F702W filter, and the spectroscopic sample is statistically
complete to $M_r\sim-19.0+5\log{h}$, within an 11\arcmin\
($\sim 1.8$ h$^{-1}$ Mpc) field.  The clusters are dynamically well-separated from the
surrounding field and most have velocity distributions consistent with Gaussians.  The velocity
dispersions range from $\sim 350$--$850$ km s$^{-1}$, consistent with the local $L_X-\sigma$ correlation.
All ten clusters host a bright, giant elliptical galaxy without emission lines, near the centre of the X-ray emission.
We measure the equivalent width
of two nebular emission lines, \oii\ and \ha, and the \hd\ absorption line to spectrally classify
the cluster members.  Galaxy morphologies are
measured from the {\it HST} images, using the two-dimensional surface-brightness fitting software {\sc gim2d}.
Emission line galaxies in these clusters are relatively rare, comprising only 22$\pm 4$\% of the sample.  
There is no evidence that these emission-line
galaxies are dynamically distinct from the majority of the cluster population, though our sample is
too small to rule out the $\sim 30$\% difference that has been observed in more massive clusters.  
We find eleven
galaxies, comprising 6\% of the cluster members, which are disk-dominated but show no sign of emission
in their spectrum.  Most of these are relatively isolated,
spiral galaxies with smooth disks. We find no cluster members with a starburst or
post-starburst spectrum.  The striking similarity between the spectral and morphological properties of
galaxies in these clusters and those of galaxies
in more massive systems at similar redshifts implies that the physical processes responsible for
truncating star formation in galaxies are not restricted to the rare, rich cluster environment, but are 
viable in much more common environments.
In particular, we conclude that ram pressure stripping
or cluster-induced starbursts cannot be solely responsible for the low star formation rates in these
systems.  
\end{abstract}
\section{Introduction}\label{sec-intro}
The evolutionary history of galaxies depends both on cosmic time and on the
type of environment in which they exist.
For example, recent studies have shown that the universal average star formation rate (SFR) has
been declining steadily since at least $z\sim 1$ \citep{L96,Madau,Cowie+99,Wilson+02}.
In a somewhat analogous way, star formation rates are known to monotonically decrease with
increasing density at a given epoch \citep[e.g.][]{B+97,P+99,2dF-sfr}.  In both
cases, the reason for the decrease in star formation is unknown.  In particular,
there is plenty of gas still available for star formation at the present day, so the
sharp decline in activity over the past $\sim 5$ Gyr is a critical issue.  
It is an intriguing possibility that the processes which influence
the evolution of galaxies that end up in dense clusters may be more generally important to galaxies in other,
more common, environments.  If this is so, it may be possible to link the 
decline in the universal average star formation rate  to environmental
effects in a Universe which is growing hierarchically.  

We now have a good empirical description for the
galaxy populations of massive clusters.  Galaxies within the virial radius have, on average,
lower star formation rates, and less recent
($<1$Gyr) star formation, than galaxies in the surrounding field;  this
is true both locally \citep{2dF-sfr,Sloan_sfr} and at higher redshifts
\citep[e.g.][]{CS87,B+97,PSG,P+99,PLO,C+01,A1689}.  Evidence is mounting that
this deficiency in star formation activity is at least partially independent
of the morphology-density relation \citep{D+97,B+98,P+99,C+01,2dF-sfr,Sloan_sfr}.
However, an explanation for this difference between cluster and field galaxies
is still lacking.  Ram pressure stripping of cold gas in the disk of
a galaxy  \citep{GG,F98,QMB} is only likely to take place in the dense cores of rich clusters,
and it seems unlikely that it can explain the suppression of star formation
as far as several Mpc from the centre \citep{B+97,Kodama_cl0939}.  Galaxy harassment \citep{harass}
may be effective at destroying small galaxy disks, but the effect that this
will have on the star formation rate of the galaxy is not clear.  On the other hand,
the observed radial and density dependences of galaxy stellar populations
and morphologies are reproduced quite well by hierarchical models in
which only the diffuse, hot halo gas expected to surround isolated galaxies is
stripped \citep{LTC,infall,Diaferio,Okamoto,Bekki02}.  The observed increase in activity with redshift
is most likely a consequence of the higher infall rate \citep{PSext,Erica,KB01}, though projection
effects may still play a role \citep{Diaferio}.

For most cluster galaxies, the last episode of star formation occurred 
many billions of years ago.  Thus, if some physical mechanism is responsible
for the transition from a more active state, it must have occurred in the
distant past, and will be difficult to uncover by observing galaxies in their
present state \citep[e.g.][]{TragerVI,deJ+D,K+01}.  It is therefore necessary to consider how the galaxy populations
have evolved over time \citep{BO84,D+99,KS00,KB01}.  However, in
hierarchical models of galaxy formation, the progenitors
of today's most massive clusters are expected to be numerous smaller
structures at higher redshifts \citep[e.g., ][]{K96}.  Thus, galaxies must
be observed not only over a range of redshifts, but for a range of cluster masses
as well.

For this purpose we have carried out an extensive observational campaign
to obtain {\it Hubble Space Telescope (HST)} imaging and ground-based imaging and spectroscopy for
ten clusters at $z\approx0.25$, selected to have low X-ray luminosities (hereafter
referred to as the \lowlx\ sample).
This sample can be directly compared with studies of more massive clusters, both
locally \citep{2dF-sfr,Sloan_sfr} and at higher redshift \citep{B+97,GPS2}.
For example, in \citet[][hereafter Paper I]{lowlx-morph} we presented an analysis
of the {\it HST} data for the present cluster sample, and compared it with a
similar {\it HST} sample of high X-ray luminosity cluster cores.  We found 
marginal ($\sim 2\sigma$) evidence
that the \lowlx\ clusters have more disk-dominated galaxies at
a fixed local density.  This suggests that at least galaxy morphology is sensitive
to the large-scale structure.  In the present paper we will revisit this,
and other issues, in light of the spectroscopic data.

The paper is organized as follows.  In \S~\ref{sec-obs} we present the
cluster selection, and the data acquisition, reduction and analysis.
Our results are presented in \S~\ref{sec-results}, where we consider
the dynamics and spectral properties of the galaxy population.
In \S~\ref{sec-discuss} we compare our results with those found for
more massive clusters, and consider the implications of
these results for models of galaxy evolution.  We summarize our
findings in \S~\ref{sec-conc}.
We use a cosmology with $\Lambda=0.7$,
$\Omega_m=0.3$, and parametrise the Hubble constant as
$H_\circ=100h$ km s$^{-1}$ Mpc$^{-1}$.

\section{Observations, Reduction and Analysis}\label{sec-obs}
%
%
{\scriptsize
\begin{table*} 
\begin{center} 
\caption{\centerline {\sc \label{tab-props} Properties of the ten clusters}} 
\vspace{0.1cm}
\begin{tabular}{lccccccc} 
\hline\hline
\noalign{\smallskip}
Name      &   R.A.\   & Dec.\                  & $N_{\rm memb}$ & $<z>$ &$\sigma$ &$L_{\rm X}$ (0.1--2.4 keV)&Completeness$^{a}$\cr
          & \multispan2{\hfil (J2000)\hfil }   &                &       &(km/s)   &$10^{43} h^{-2}$ ergs s$^{-1}$&\cr 
\noalign{\smallskip}
\hline
\noalign{\smallskip}
Cl\,0818+56 & 08 19 04 & +56 54 49 & 9 &   0.2670    & 651$\pm165$     & 1.50&0.66\cr
Cl\,0819+70 & 08 19 18 & +70 55 48 & 23 &  0.2296    & 356$\pm39$      & 1.26&0.88\cr
Cl\,0841+70 & 08 41 44 & +70 46 53 & 21 &  0.2397    & 399$\pm170$      & 1.22&0.52\cr
Cl\,0849+37 & 08 49 11 & +37 31 09 & 26 &  0.2343    & 764$\pm90$      & 1.93&0.69 \cr
Cl\,1309+32 & 13 09 56 & +32 22 14 & 19 &  0.2932    & 662$\pm1304$     & 2.01&0.41\cr
Cl\,1444+63a & 14 43 55 & +63 45 35 & 13 & 0.2923    & 403$\pm73$      & 3.99$^b$&0.49\cr
Cl\,1444+63b & 14 44 07 & +63 44 59 & 15 & 0.3006    & 449$\pm681$      & 3.99$^b$&0.49\cr
Cl\,1633+57 & 16 33 42 & +57 14 12 & 18 &  0.2402    & 582$\pm360$      & 0.49&0.87\cr
Cl\,1701+64 & 17 01 47 & +64 20 57 & 12 &  0.2458    & 834$\pm647$       &0.40&0.52\cr
Cl\,1702+64 & 17 02 14 & +64 19 53 & 15 &  0.2233    & 386$\pm426$      & 0.74&0.52\cr
\noalign{\hrule}
\end{tabular}
\end{center} 
\begin{flushleft}
$^{a}$Computed for galaxies more than 1 magnitude brighter than the faintest galaxy with a redshift in that cluster.\\
$^{b}$Cl\ 1444+63 is only detected as a single X-ray source; this $L_{\rm X}$ presumably includes contribution from both clusters. \\
\end{flushleft}
\end{table*}
}
\subsection{Cluster Selection}\label{sec-sample}
The cluster sample is the same as the \lowlx\ sample analysed in
Paper~I, with the addition of Cl\,1633+57, comprising
ten X-ray faint clusters in the northern hemisphere\footnote{The cluster Cl\,1444+63 turns out to
be two clusters aligned along the line of sight (see \S~\ref{sec-redux}), and
we consider the two clusters separately.}.
The clusters are selected from the
sample identified by \citet{V+98}
in serendipitous, pointed {\it ROSAT} {\it PSPC} observations,
restricted to a
relatively narrow redshift range, $z=0.22$--0.29 ($\sigma z/z\sim 0.1$)
and a mean redshift of $z=0.25$, to reduce the effects of differential
distance modulus and k-correction effects on the comparison between
the systems.  
We compute $L_{\rm X}$ in the 0.1--2.4 keV band from the observed
fluxes in the 0.5--2.0 keV band, corrected for galactic H{\sc i} absorption and assuming a k-correction
appropriate for an intra-cluster gas temperature equal to that expected from
the local $L_{\rm X}-kT$ relation \citep{AF,Markevitch},
using the software package {\sc xspec}.  These luminosities are listed in Table~\ref{tab-props}, and
range from $0.40$ to $4.0\times 10^{43} h^{-2}$ ergs s$^{-1}$[0.1--2.4 keV] (Table~\ref{tab-props}). 
Although Cl\,1444+63 is treated as two separate clusters, the X-ray luminosity in Table~\ref{tab-props} is that
of the combined clusters.

\subsection{Imaging and Spectroscopy}\label{sec-imspec}
{\it HST}\, imaging with {\it WFPC2} is available
for all clusters in the sample.
The observations of all clusters but Cl\,1633+57 are described more completely in Paper~I.
To summarize, each cluster was observed with three single orbit exposures
in the F702W filter during Cycle~8.  The photometry is calibrated on the Vega
system, with updated zero
points taken from the current instrument manual.  The final images
reach a 3-$\sigma$ point source sensitivity of $R_{702} \sim 25.5$, and
cover a field of $2.5' \times 2.5'$ (or $0.4h^{-1}$\,Mpc at $z=0.25$)
with an angular resolution of 0.17$''$ ($\sim$0.5$h^{-1}$\,kpc).  
The cluster Cl\,1633+57 was observed in Cycle 8 (Proposal ID 7374), and the data 
were retrieved from the CADC\footnote{Canadian Astronomy Data Centre, 
which is operated by the Herzberg Institute of Astrophysics, National Research Council of Canada.} 
{\it HST} archive.  These data are F702W WFPC2 observations, with four
exposures of 1200 seconds.  The total exposure time of 4800 seconds
is therefore less than that of the other nine clusters (typically
7000 seconds; see Paper~I).  The calibrated images obtained from the
CADC were combined in the same way as described in Paper~I.

The spectroscopic sample is selected from ground-based imaging
from the Palomar 200-inch telescope and the Isaac Newton telescope
(INT) .   The single INT Wide-Field Camera (WFC) chip from which the galaxies were
selected covers 11.4\arcmin\ at  0\farcs33 arcsec pix$^{-1}$, while
the Palomar COSMIC images have a field-of-view of
13.7\arcmin with a pixel scale of 0\farcs4 arcsec pix$^{-1}$ \citep{COSMIC}.
For all but two clusters, galaxies were selected for spectroscopic follow-up
from the $R-$band images.  In Cl\,1309+32 and Cl\,1444+63 the
sample was selected  from $I-$band images, because $R$ was not available.
Note that the two
clusters Cl\,1701+64 and Cl\, 1702+64 are sufficiently close together that spectroscopic
targets for both could be selected from a single WFC chip.
The conditions during the imaging observations were not photometric,
so we have calibrated our images by comparing aperture magnitudes
of several (usually 2-3) relatively isolated, early-type galaxies with the F702W photometry of the {\it WFPC2} images,
and converted this to standard $R$ magnitudes, assuming $R_{\rm F702W}-R_c=-0.2$ \citep{F+95}.
Because of uncertainties in the colour term, and the small number of calibration galaxies
used for each cluster, the photometric calibration is likely to be accurate to
only $\sim 0.2$ mag.  

The spectra were obtained over four observing runs, and we give a log of these
in Table~\ref{tab-obs}.  Three sets of
data were taken  with MOSCA on
the 3.5-m telescope at Calar Alto Observatory, using the g500 grism.
The spectra have a dispersion of $\sim 2.7$\AA~pix$^{-1}$ and cover
4000\AA$-$8000\AA, with a resolution of $\sim 10-15$\AA. 
The fourth observing run was with 
LDSS-2 on the William
Herschel Telescope.  Using the medium-blue grism, we obtained a
dispersion of $\sim 4.5$\AA~pix$^{-1}$, covering 3500--9000\AA.
The resolution of these spectra is $\sim 15$--$20$\AA.  
Typically, two masks were observed for each cluster;
in some cases a third mask was also obtained.  
Galaxies were selected for spectroscopic follow up based solely on their
instrumental $R$ or $I$ band magnitude, with preference given to brighter galaxies.
The fraction of galaxies observed spectroscopically therefore declines toward
fainter magnitudes.  
For each mask we obtain between
20 and 35 spectra through 1\farcs5 wide slits,  over an 11\arcmin\ field of
view.  In total we
obtained 581 spectra, in variable conditions.  Some galaxies
observed in poor conditions were later reobserved in a subsequent run.
We obtained reliable redshifts for a total of 317 galaxies, of which 172 are
cluster members.  A summary of the photometric and spectroscopic observations is given in
Table~\ref{tab-obs}.
%
%
{\scriptsize
\begin{table} 
\begin{center} 
\caption{\centerline {\sc \label{tab-obs} Log of Observations}} 
\vspace{0.1cm}
\begin{tabular}{lllcc} 
\hline\hline
\noalign{\smallskip}
Name      & Instrument &   Date     &  Band (phot)                           &$T_{\rm exp}$\cr
          &            &               &  $N_{\rm mask}$/ $N_{\rm spec}$ & (ks)        \cr 
\noalign{\smallskip}
\hline
\noalign{\smallskip}
Cl\,0818+56 & COSMIC & 26/11/98   & R      & 0.3     \cr
  & MOSCA       & 04/02/00    & 2 / 53 & 7.2     \cr
\smallskip  & LDSS2       & 02/03/00  & 2 / 53 & 5.4/7.2 \cr 
Cl\,0819+70 & COSMIC & 26/11/98   & R      & 0.25    \cr
            & MOSCA       & 03/11/00    & 2 / 39 &  7.2    \cr
\smallskip  & LDSS2       & 04/03/00  & 1 / 21 & 7.2     \cr 
Cl\,0841+70 & COSMIC & 26/11/98   & R      & 0.25    \cr
            & MOSCA       & 04/02/00    & 2 / 31 &  7.2    \cr
\smallskip  & LDSS2       & 05/03/00  & 1 / 21 & 7.2     \cr
Cl\,0849+37 & COSMIC & 26/11/98   & R      & 0.25    \cr
            & MOSCA       & 14/04/99 & 1 / 29 &  5.4    \cr
\smallskip  & MOSCA       & 05/02/00    & 2 / 41 & 7.2/9.6 \cr
Cl\,1309+32 & INT/WFC     & 19/06/98  & I      & 0.6     \cr
            & MOSCA       & 12/04/99 & 2 / 89 & 7.2/9.0 \cr
\smallskip  & MOSCA       & 05/02/00    & 1 / 31 &  9.6    \cr
Cl\,1444+63 & INT/WFC     & 18/01/99   & I      & 0.72    \cr
            & MOSCA       & 13/04/99 & 1 / 31 &  9.0    \cr
\smallskip  & MOSCA       & 06/02/00    & 1 / 36 &  7.2    \cr
Cl\,1633+57 & INT/WFC     & 10/02/99   & R      & 0.6     \cr 
\smallskip  & MOSCA       & 12/04/99 & 2 / 58 &  7.2    \cr
Cl\,1701+64/&INT/WFC&10/02/99& R     & 0.6     \cr
Cl\,1702+64  & MOSCA       & 30/07/00  & 2 / 48 & 5.4     \cr
\noalign{\hrule}
\end{tabular}
\end{center} 
\end{table}
}
\subsection{Data Reduction and Analysis}\label{sec-redux}
The spectroscopic data were reduced using {\sc iraf}\footnote{{\sc iraf} is distributed by the National Optical Astronomy Observatories which is 
operated by AURA 
Inc. under contract with NSF.} software.  The images
were bias-subtracted and median-combined to remove cosmic rays.
Spectra were optimally extracted and the sky was subtracted by fitting 
a one- or two-degree polynomial to
the counts on either side of the object.  The spectrum was traced along the
dispersion direction of the CCD to account for distortion.  Wavelength calibration was based
on either HgNeAr arc lamps or emission lines in the night-sky spectrum; the latter
method was generally used to improve the calibration at the red end of the spectrum.
The {\it r.m.s.} of the wavelength solution is typically $\lesssim 1$\AA\ for $\lambda\lesssim 7000$\AA,
corresponding to $\Delta z \lesssim 2\times 10^{-4}$.    
The spectra were not flat-fielded or flux calibrated, which is inconsequential for this
analysis since we restrict our attention to spectral features that are defined
relative to the continuum, over a narrow
wavelength range ($\sim 150$\AA).

Galaxy redshifts were determined by three of us (MLB, BLZ, AF), each with an
independent method.  MLB used the routine {\sc fxcor} within the {\sc iraf}
environment to cross-correlate the spectra with high signal-to-noise $z=0$
galaxy spectral templates of similar resolution.  BLZ used the Fourier
Correlation Quotient \citep{FCQ} within {\sc midas}, which compares absorption lines
with those of template stars,
while AF measured the centroids of several prominent absorption lines.
The agreement between the three measurements is good, and generally within
the uncertainty (typically $\sim 100$ km s$^{-1}$).  The independent estimates
allowed us to identify galaxies where one method failed (because of low signal-to-noise
ratio around a critical line, for example).  Spectra for which we could not
resolve the discrepancy between different redshift estimates ($\lesssim 5\%$)
were always of low signal-to-noise ratio, and were removed from further analysis.

\begin{figure}
\leavevmode \epsfysize=8cm \epsfbox{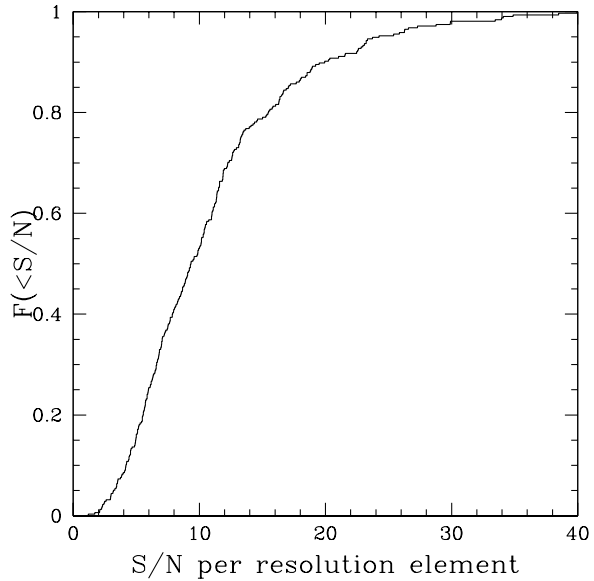}
\caption{The cumulative distribution of signal-to-noise ratio per resolution element,
for the 315 galaxies in our sample with redshifts.  The resolution element
is $\sim 15$\AA\ for the MOSCA spectra, and $\sim 20$\AA\ for the LDSS-2
spectra.
\label{fig-snr}}
\end{figure}

In Figure ~\ref{fig-snr} we show the distribution of signal-to-noise
ratio per resolution element, including only those spectra for which a redshift was
obtained.  The S/N is computed in the rest-frame continuum region
$4050$--$4250$\AA, redward of the 4000\AA\ break, and has a median of $\sim 10$ per
resolution element.  It can easily be shown that, for spectra with a resolution of $15$ \AA,
an emission line can be measured with an uncertainty $<5$ \AA\ if $S/N > \sqrt{18+(W/5)^2}$,
where $W$ is the equivalent width of the line, assuming the uncertainty is dominated by the
continuum flux (true for weak lines).  Thus, lines as weak as $W\sim 5$\AA\ can
be reliably measured if $S/N>4.3$\AA, which is satisfied by $\sim$90\% of our spectra.

\subsection{Selection Function and Magnitude Limits}\label{sec-selfun}
\begin{figure}
\leavevmode \epsfysize=8cm \epsfbox{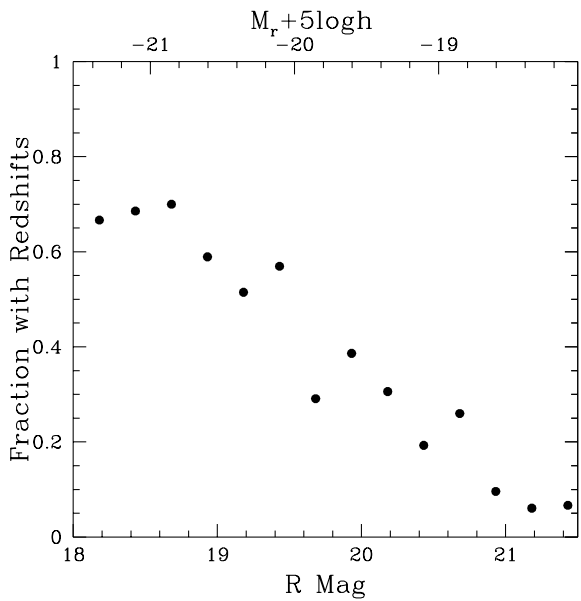}
\caption{The average selection function for the whole sample.   The top axis shows the k-corrected
luminosity for a cluster at z=0.25.
In practice,
the data are weighted by the selection function, as a function of instrumental
magnitude, for each cluster separately. 
\label{fig-selfun}}
\end{figure}

For each cluster, we determine the fraction of galaxies
in the photometric catalogue for which a redshift was obtained, as a function
of instrumental magnitude.  
Galaxies in our spectroscopic sample are then weighted by the inverse of this selection function to obtain a 
sample which is statistically magnitude limited.  The selection function
is determined separately for each cluster, as a function of instrumental
magnitude.  In Figure~\ref{fig-selfun} we show
the combined selection function for the full sample, as a function of $R-$magnitude.
We define the completeness as the fraction of galaxies with a redshift
in each cluster field, considering
only galaxies brighter than 1 magnitude above the faintest galaxy
for which a redshift was obtained.  This quantity is tabulated in Table~\ref{tab-props}.

Our sample is $\gtrsim 20\%$ complete at $R\sim 20.5$, which corresponds to
$M_R\sim -19.0+5\log{h}$ at $z=0.25$ (including a k-correction of 0.2 mag).
Unless otherwise stated, we limit our sample to galaxies brighter than this magnitude, which
corresponds to $\sim$2 magnitudes fainter than $M^\ast$.  

\subsection{Line Index Measurements}\label{sec-indices}
\begin{figure}
\leavevmode \epsfysize=8cm \epsfbox{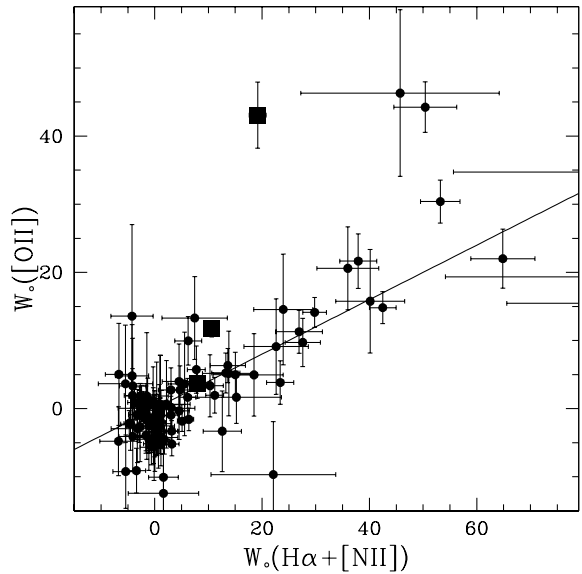}
\caption{The correlation between \ewha\ and \ewoii\ for galaxies in
which a reliable measurement of both lines exists.  Error bars are
1$-\sigma$.  The three galaxies represented by {\it filled squares} 
have strong, broad [N{\sc ii}] emission characteristic of non-thermal
emission.  The solid line is the average local relation measured
by \citet{K92}.  Three points are off the scale, at \ewha$>80$\AA;
part of the error bars are just visible.  
\label{fig-oiiha}}
\end{figure}
We have shown that the signal-to-noise ratio and resolution of our spectra are 
sufficient to allow us to reliably measure spectral features with equivalent
widths $W \gtrsim  5$\AA.
In particular we focus on the
rest frame equivalent width of the \oiifull\ emission line, which
is a signature of gas ionized
by hot stars and, hence, the best indicator of current star formation 
in the blue region of the spectrum \citep[e.g.,][]{K83}.  However,
it is less than perfect for several reasons.  Perhaps most
importantly, it is a weak line, usually with an equivalent width $<30$\AA.  In moderate signal-to-noise
spectra like ours, therefore, it is difficult to detect low levels of
star formation.  Furthermore, the equivalent width of \oii\ is sensitive
to metallicity and the ionization level of the gas \citep{CL,CL_ext}, which
can introduce a scatter in the \oii-SFR relation of a factor $\gtrsim 5$.
Finally, it seems likely that the H{\sc ii} regions where \oii\ is produced
will be more heavily obscured than the longer-lived stars which give rise
to the continuum \citep[e.g.][]{Silva,CF00}.  In this case, the equivalent width
will underestimate the star formation rate, even if a global extinction
correction is applied.  However, despite these difficulties, it has been
shown that \oii\ correlates with better star formation indicators like
H$\alpha$ emission \citep[also see Fig~\ref{fig-oiiha}]{K92,Jansen}.  Although there is a large scatter in this correlation,
the average star-formation properties of a population can be
well-described.  

We also measure the H$\delta$ absorption line, which is a strong feature
in A-stars and thus represents recent star formation.  As this is an
absorption feature, and affected by emission-filling, its precise measurement is
more difficult than that of the emission lines, and little can be said
about most of the galaxy population.  However, galaxies with very strong
H$\delta$ absorption are easily detected, and these galaxies may play
an important role in galaxy evolution \citep{DG83,CS87,P+99,PSG}.

These rest-frame equivalent widths are measured by fitting a line
to the continuum on either side of the feature, and summing the flux above
this continuum over a well-defined wavelength range.  The definitions of the
\ewoii\ and \ewhd\ indices are the same as in \citet{PSG}\footnote{Note that
the \ewhd\ feature wavelength range given in Table~1 of \citet{PSG} is incorrect.  In both
that work and the present analysis, the feature is taken to cover $4088$--$4116$\AA.}.  
We will adopt the
convention that \ewoii\ is positive when in emission, and \ewhd\ is
positive in absorption.  
Each line measured is checked to ensure that the continuum fit or
spectral line pixels are not adversely affected by bad sky subtraction, bad pixels, or poor wavelength
calibration.  This removes 7.8\% of
the cluster member \oii\ measurements from our analysis.   Uncertainties
are computed as described in \citet{PSG}, and include the Poisson noise contributed
from the sky subtraction and, therefore, the degradation effected by the
spectral resolution. 

There are 167 cluster members with reliable measurements of \oii, and 107 of these 
also have reliable measurements (not necessarily detections) of H$\alpha$+[N{\sc ii}].  Thus,
we can use measurements of this line to check the reliability of
\oii\ as a star formation indicator in this sample.  We measure \ewha\ in 
a manner analogous to that for \ewoii, by computing the flux in the range $6555$\AA$<\lambda<6575$\AA, compared with
the continuum in the regions $6490$\AA$<\lambda<6537$\AA\ and $6594$\AA$<\lambda<6640$\AA.
We do not deblend the two adjacent [N{\sc ii}] lines, but include them both in the equivalent width
measurement. The [N{\sc ii}]$\lambda6548$
line always contributes only a small amount of flux ($<5$\%), while the stronger
[N{\sc ii}]$\lambda6583$ line contributes $\sim 30$\% of the
flux. 
The observed correlation between \ewha\ and \ewoii\ is in excellent agreement with the local field correlation
of \citet{K92}, as shown in Figure~\ref{fig-oiiha}.  
Galaxies for which \ewoii$>5$\AA\ are almost always detected with \ewha$>20$\AA. 
Only three ($4$\%) of these galaxies 
have strong, broad [N{\sc ii}]$\lambda6583$ indicating that their emission is
dominated by non-thermal processes \citep{VO87,K+01}.

\subsection{Morphologies}\label{sec-morph}
{\scriptsize
\begin{table} 
\begin{center} 
\caption{\centerline {\sc \label{tab-morphs} Fraction of Disk Galaxies}} 
\vspace{0.1cm}
\begin{tabular}{lcccc} 
\hline\hline
\noalign{\smallskip}
Cluster      & $F_{\rm stat}$ ($R<23$)& $F_{\rm stat}$ ($R<20$)& $F_{\rm spec}$\cr
             & (Paper~I)              &                        &          \cr 
\noalign{\smallskip}
\hline
\noalign{\smallskip}
Cl\,0818+56 & 0.43$\pm$0.08&0.64$\pm$0.17&0.56$\pm$0.38\cr
Cl\,0819+70 & 0.15$\pm$0.37&0.07$\pm$0.23&0.00$\pm$0.10\cr
Cl\,0841+70 & 0.36$\pm$0.27&0.04$\pm$0.12&0.00$\pm$0.05\cr
Cl\,0849+37 & 0.45$\pm$0.10&0.22$\pm$0.12&0.14$\pm$0.17\cr
Cl\,1309+32 & 0.37$\pm$0.09&0.36$\pm$0.15&0.14$\pm$0.17\cr
Cl\,1444+63 & 0.23$\pm$0.10&0.20$\pm$0.12&0.20$\pm$0.20\cr
Cl\,1633+57 & 0.26$\pm$0.18&0.17$\pm$0.18&0.00$\pm$0.10\cr
Cl\,1701+64 & 0.50$\pm$0.10&0.36$\pm$0.16&0.32$\pm$0.25\cr
Cl\,1702+64 & 0.47$\pm$0.12&0.42$\pm$0.19&0.00$\pm$0.14\cr
\noalign{\hrule}
\end{tabular}
\end{center} 
\end{table}
}
There are 78 galaxies for which we have both spectroscopy and {\it HST}
{\it WFPC2} imaging; 62 of these are confirmed cluster members
(see \S~\ref{sec-dynamics} for the membership definition).
As described in Paper~I, we have measured fractional bulge
luminosities, B/T, from these images using the two-dimensional surface
brightness-fitting software package {\sc gim2d} \citep{Gim2d}.
The {\it HST} imaging only covers the 
central $\sim 0.6$ Mpc in these clusters, and the number of disk-dominated,
confirmed cluster members is small: only eight galaxies ($18 \pm 5$\%) brighter than $R=20$ have $B/T<0.4$. 
This is much less than the $\sim 40\%$ fraction of disk-dominated galaxies found in Paper~I.
This difference can be attributed to the different limiting magnitudes of the two
analyses.  The present spectroscopic sample is limited to $R\sim 20$, while in
Paper~I we considered a sample three magnitudes fainter, using
only the {\it HST} imaging data.   In Table~\ref{tab-morphs} we list as $F_{\rm stat}$ the fraction of disk galaxies
($B/T<0.4$) for each cluster, with a statistical background subtraction
based on the Medium Deep Survey \citep[][see Paper I for details]{MDS-data,MDS-morph3}.
We show both the value computed to $R\sim 23$, as published in Paper~I, and
the value considering only galaxies with $R<20$.  These numbers are compared
with $F_{\rm spec}$, the fraction of disk-dominated galaxies determined from
the spectroscopic sample of cluster members, complete to $R\lesssim 20$ (we combine the two clusters
Cl\, 1444a and Cl\,1444b to be consistent with the measurement of $F_{\rm stat}$).  
The errors are all computed from
bootstrap resampling, except in the case where no disk galaxies are found, where
we assume a Poisson distribution.
Adopting the brighter limit appropriate for our
spectroscopic sample, the fraction of disks determined by photometric
field correction is 26$\pm 17$\%.  
This number is consistent with the spectroscopically-determined number ($18 \pm 5$\%), within the
statistical uncertainties.

\subsection{The Catalogue}\label{sec-data}
A catalogue of the relevant derived quantities for all cluster
members (see definition in \S~\ref{sec-dynamics}) is given
in Table~\ref{tab-data}.   Galaxies are identified by an identification number from the
photometric catalogue in column 1; serendipitous observations not in the original catalogue
have an identification number of $-99$.  The $r$ magnitude, redshift and sky coordinates are
given in columns 2--5.  For galaxies observed with {\it HST}, column 6 lists the {\sc GIM2D} B/T measurement.
Note that the formal errors on B/T given by
{\sc GIM2D} are typically $\lesssim 0.02$; however these errors do not 
account for uncertainties in the sky level or point spread function, which
likely dominate the measurement for faint galaxies.  We have not, therefore,
listed these uncertainties in Table~\ref{tab-data}.
The spectral signal-to-noise
ratio per resolution element is given in column 7, and the equivalent widths of
\ewoii, \ewhd, and \ewha\ are listed in columns 8-10.  

In Figure~\ref{fig-normal1} we show the images
and spectra for all the cluster members with {\it HST} data. 
The brightest, central cluster members are shown separately in Figure~\ref{fig-BCG},
as are the sample of disk-dominated galaxies without emission lines (see \S~\ref{sec-anemic}),
which are shown in Figure~\ref{fig-anemic}.
Note that the spectra in the vicinity of H$\alpha$ are usually dominated by residuals
from night-sky emission lines, not real features or Poisson noise.  Also, the wavelength
solution beyond $\lambda\sim7000$ \AA\ is occasionally inaccurate, resulting in a small
misalignment with H$\alpha$.  This has a negligible effect on our measurement, since
the H$\alpha$ line still always lies within the bandpass we use to measure the flux.

\begin{figure*}
\centerline{\psfig{file=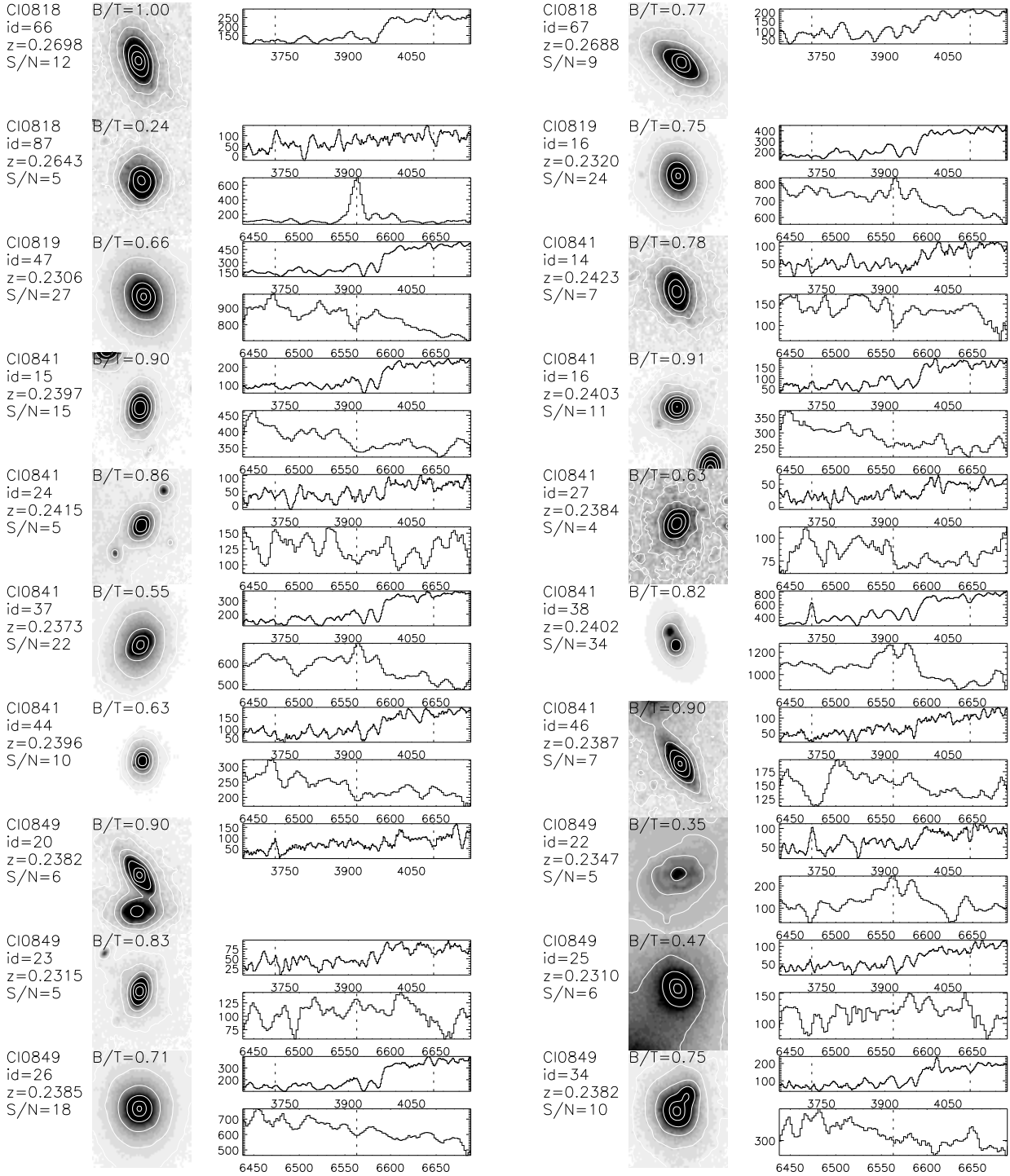}}
\caption{
Images and spectra for all cluster members observed with {\it HST}.  The
F702W images are shown with logarithmically
spaced contours over-plotted, and the B/T measurements from {\sc GIM2D} are printed on
the images.  All images are $3\arcsec\times 3\arcsec$, oriented north to the top and east
to the left.  To the left of the image,
the galaxy identification and redshift are shown, together with the signal-to-noise
ratio per resolution element of the spectrum, over $4050-4250$\AA\ (rest frame).
The spectra are smoothed to the instrumental
resolution of $\sim 15$\AA.  For every galaxy we show the region around the \oii\ and H$\delta$ lines,
with the location of those lines marked with a vertical, dotted line.  When the
spectrum near H$\alpha$ is sufficiently clear of night-sky lines to permit a reasonable
measurement of the H$\alpha$ line strength, this region of the spectrum is also shown,
with the position of H$\alpha$ indicated.  The y-axis of the spectra gives the number of
CCD counts, after sky subtraction; the x-axis is the wavelength, in \AA.
\label{fig-normal1}}
\end{figure*}\setcounter{figure}{3}
\begin{figure*}
\centerline{\psfig{file=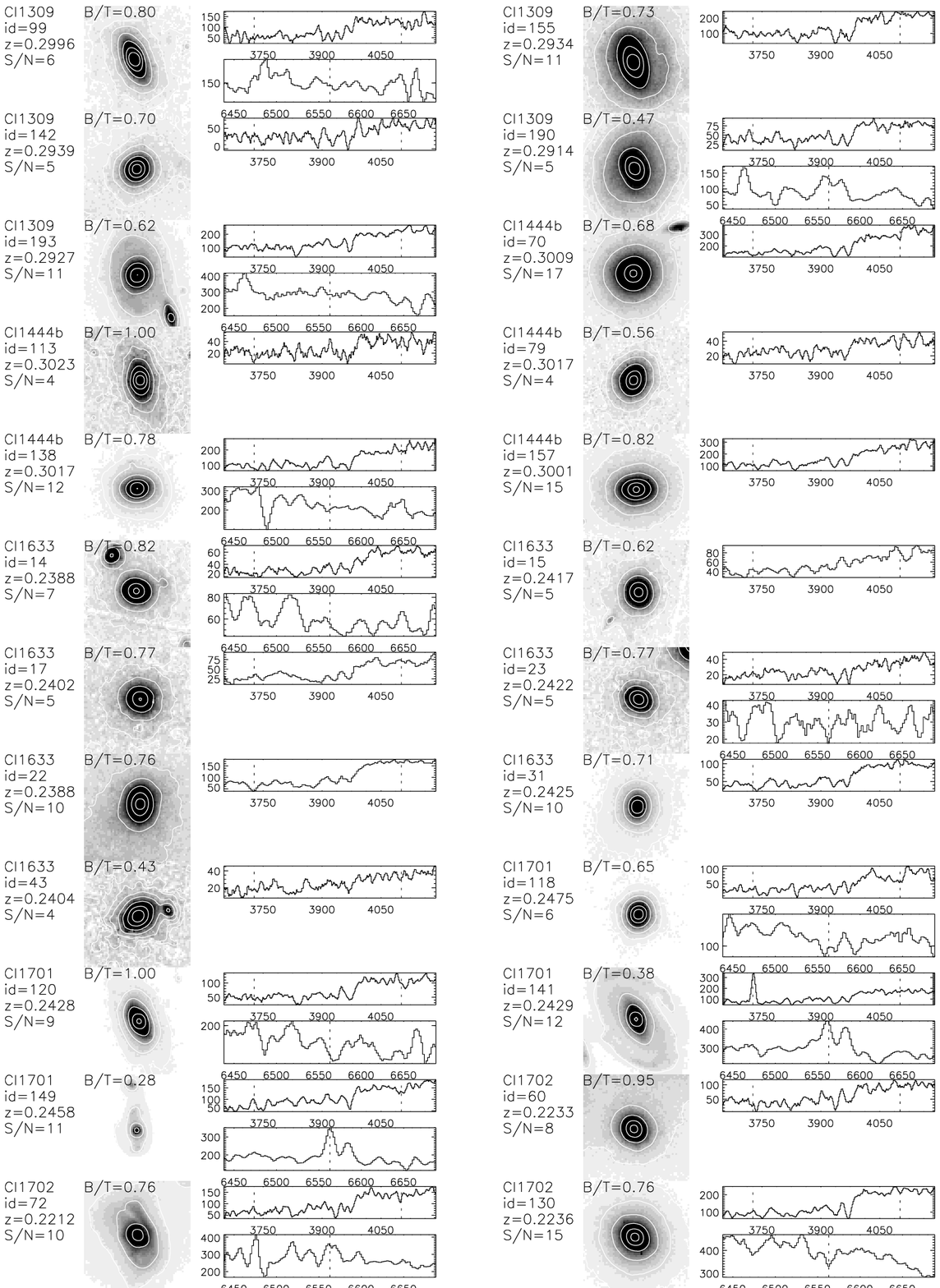}}
\caption{continued}
\end{figure*}

\begin{figure*}
\centerline{\psfig{file=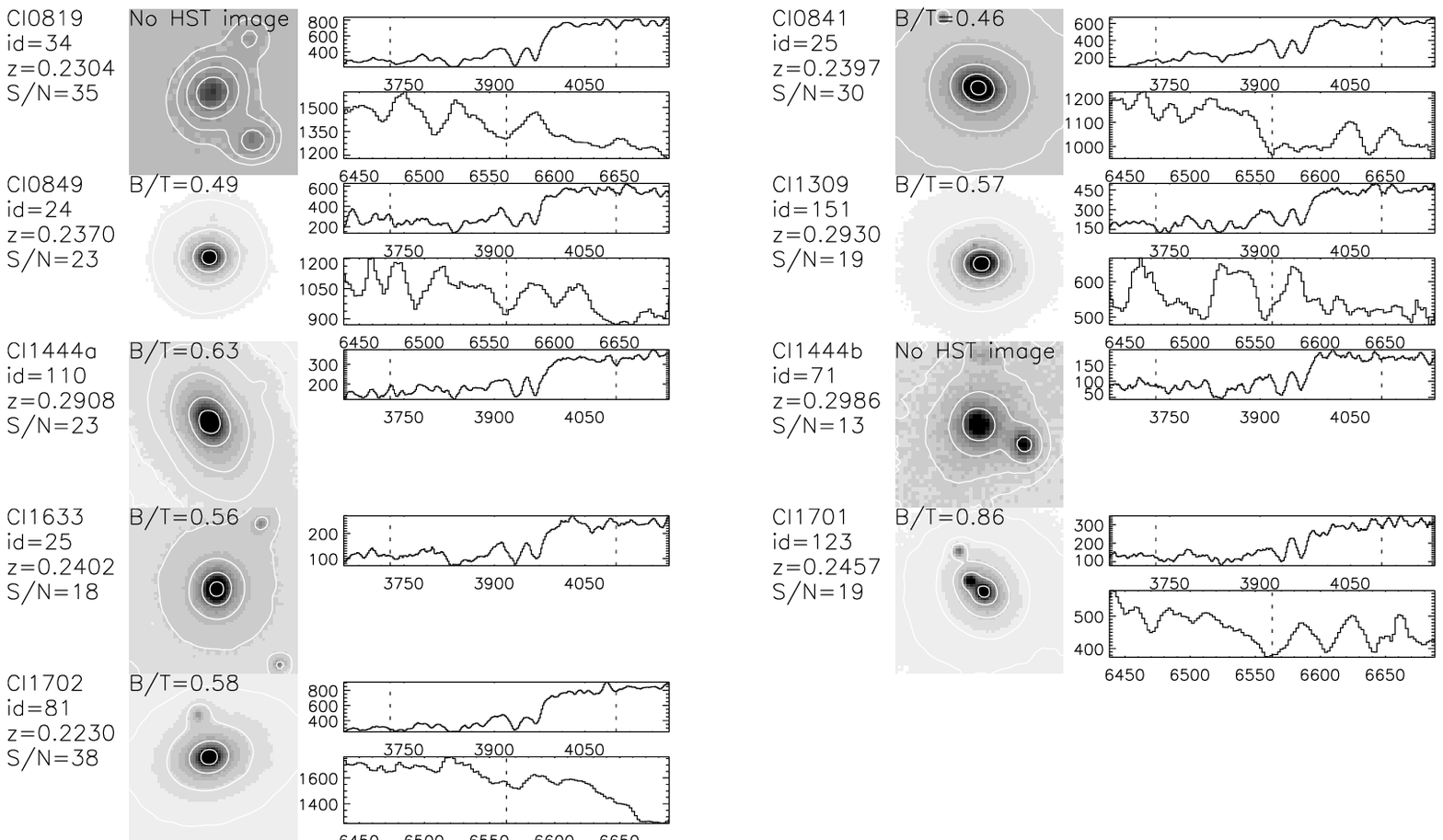,height=9.cm}}
\caption{The central, bright galaxies of each of the clusters.  The format of the figure is the same as
for Figure~\ref{fig-normal1}.  The central galaxy of Cl0818 lies behind a bright, foreground spiral which dominates the
observed spectrum, and is not shown.
Ground-based images are shown where {\it HST} images are not available; these are
also 3\arcsec\ on a side.
\label{fig-BCG}}
\end{figure*}
\begin{figure*}
\centerline{\psfig{file=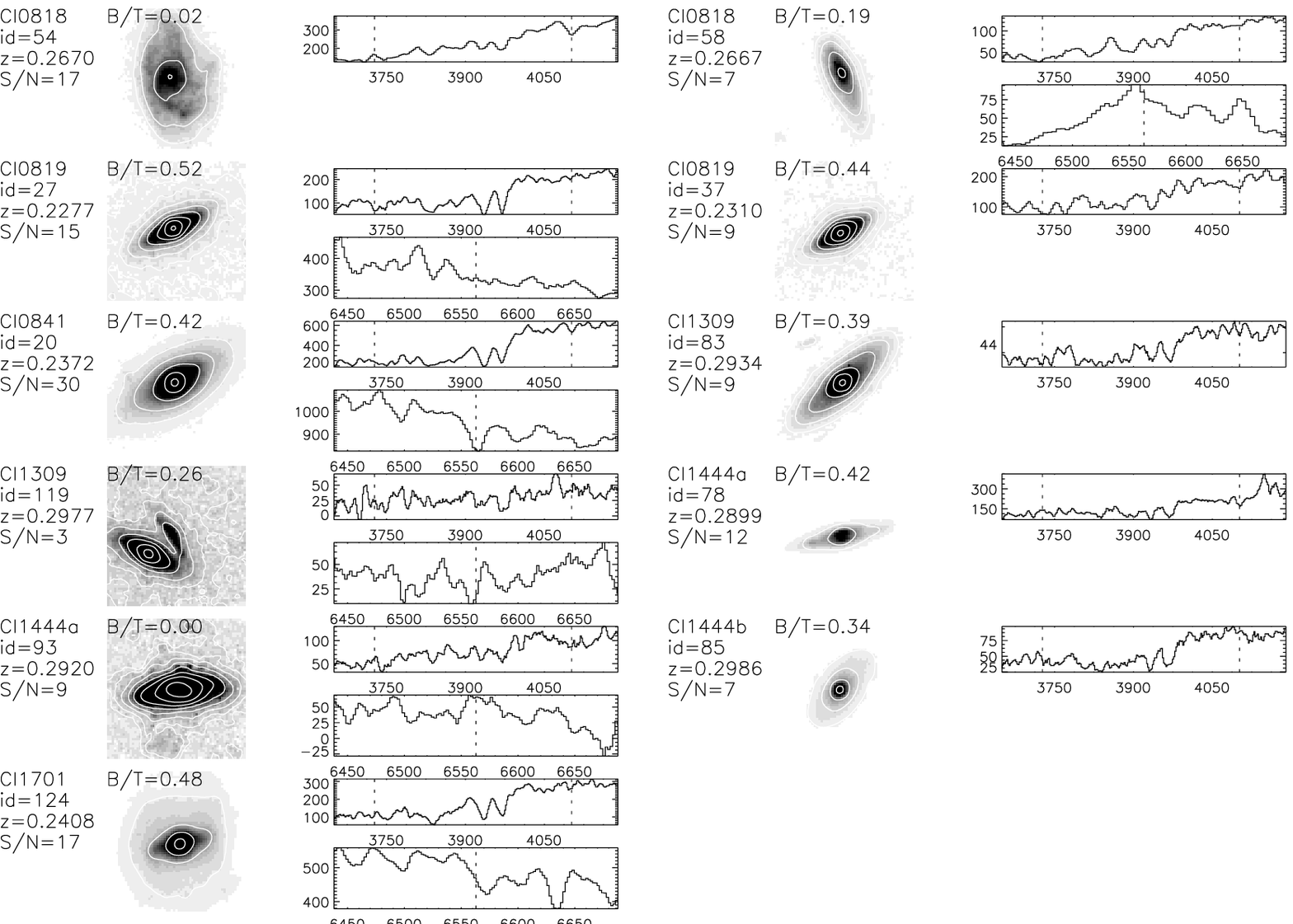,height=11.cm}}
\caption{Disk-dominated galaxies without emission lines.  The format of the figure is the same as
for Figure~\ref{fig-normal1}. 
\label{fig-anemic}}
\end{figure*}

\section{Results}\label{sec-results}
\begin{figure}
\leavevmode \epsfysize=8cm \epsfbox{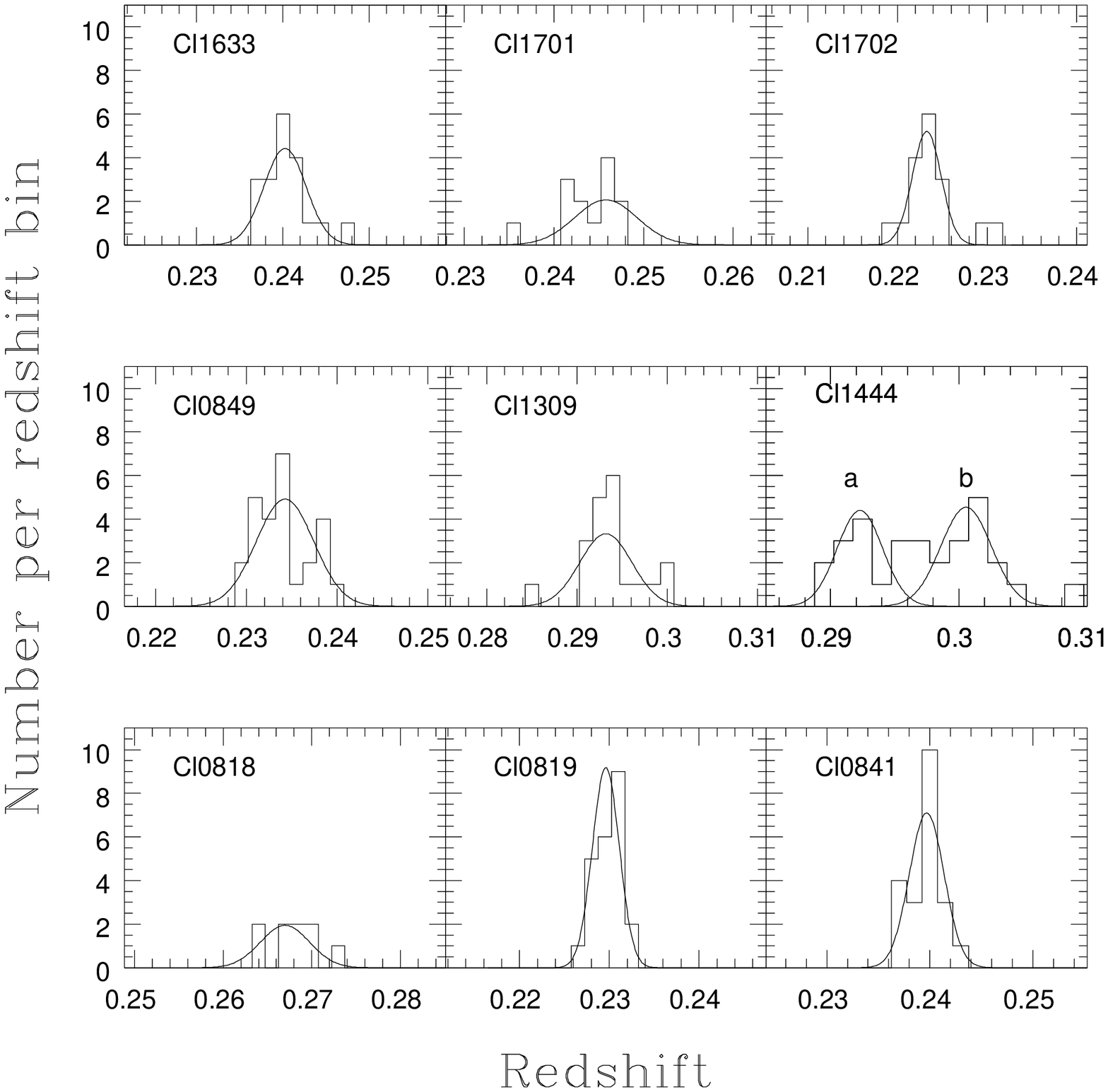}
\caption{Redshift distributions for each of the clusters in our sample.
The redshift range plotted represents 9000~km~s$^{-1}$ about the mean
redshift, in the rest-frame of the cluster.
The smooth curves overlayed are Gaussian functions with the computed centre 
and observed-frame velocity dispersion for each cluser.  Note the dispersions
tabulated in Table~\ref{tab-props} are in the rest-frame of the cluster and
include a small correction for the 100 km s$^{-1}$ uncertainty on the redshifts.
\label{fig-vdisp}}
\end{figure}
\subsection{Dynamics}\label{sec-dynamics}
In Figure \ref{fig-vdisp} we show the redshift histogram of each cluster.
For most clusters, the median redshift and its dispersion were determined using
the biweight estimator \citep{Beers} on galaxies within $\Delta z = 0.02$ of the
cluster.  The exception is the double cluster Cl\,1444, which has a bimodal redshift distribution, with
a rest-frame velocity difference of 2080 km s$^{-1}$.  In this case, the biweight
estimator results in an unreasonably high dispersion for each peak in the distribution.  
Instead, we use a 2-sigma clipping method to better characterise the distribution
of each structure. 
We recognize this treatment as arbitrary, and do not consider the dispersions
of these two clusters to be well determined.
A small correction is made for the estimated uncertainty of $100$~km~s$^{-1}$ on each redshift,
by subtracting these in quadrature from the measured dispersion.
For most of the clusters the velocity distribution includes
$\sim 20$ galaxies, and is well fit by a Gaussian, shown as the smooth
curves over-plotted in Figure \ref{fig-vdisp}.
There appears to be a structure
in redshift space between Cl\,1701 and Cl\,1702; with the small number of redshifts
available, it is not possible to determine whether or not this substructure is associated
with one of the clusters.
The remaining clusters are well isolated from the surrounding field.  Cluster
members are taken to be all those within 3$\sigma$ of the cluster velocity
dispersion.  

The velocity dispersion,
its uncertainty, and the number of galaxies within 3$\sigma$ of each cluster are
listed in Table~\ref{tab-props}.
The velocity dispersions of all the clusters
are fairly similar, with a mean $\sigma=548$
km s$^{-1}$ and a standard deviation of 172~km~s$^{-1}$.  
The uncertainty in velocity dispersion, determined by jackknife resampling, is generally quite
large, $>200$ km s$^{-1}$.  In particular, it is much larger for
the clusters Cl\,1309, Cl\,1701 and Cl\,1702, which appear to have substructure
in the wings of the distribution.
We expect the
virial radii to be $R_v$[Mpc]$\sim 0.0035(1+z)^{-1.5}\sigma$[km/s]$\sim
1.4$ Mpc \citep{Girardi98}, or $\sim 5.9$ arcmin at $z=0.25$.
Because of the large uncertainties on the
velocity dispersions, however, the virial radii of individual clusters are not
well determined.

\begin{figure}
\leavevmode \epsfysize=8cm \epsfbox{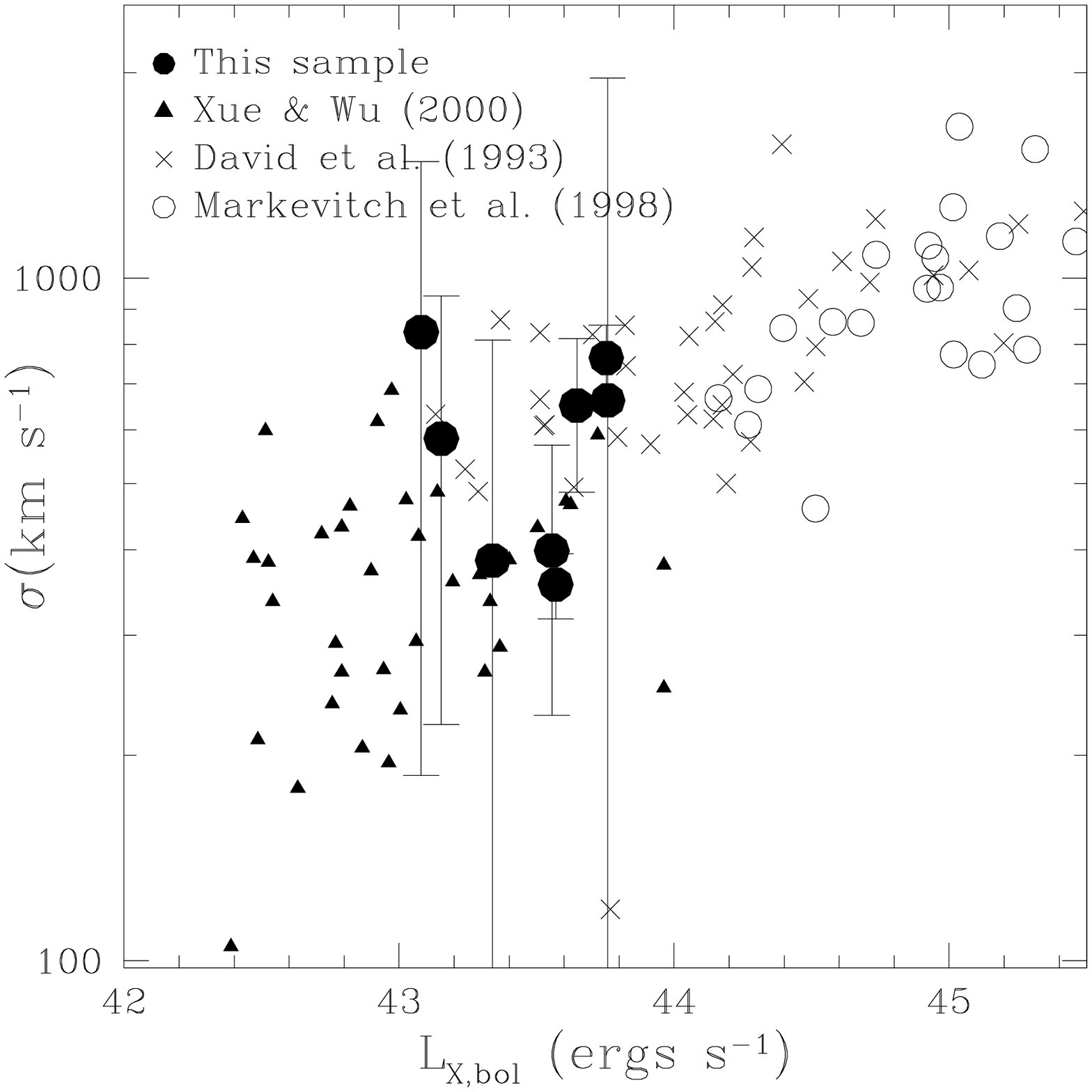}
\caption{The correlation between velocity dispersion $\sigma$ and bolometric X-ray
luminosity ($\Lambda=0.7, \Omega_m=0.3, h=0.7$) for our sample is shown as {\it solid
circles} with 1$-\sigma$ error bars on the velocity dispersion.  
This is compared with three local cluster samples, as indicated; error bars are omitted
on these points for clarity.
\label{fig-Lsig}}
\end{figure}

The measured velocity dispersions are in good agreement with those expected from
the local correlation with X-ray luminosity, as shown in Figure~\ref{fig-Lsig}.
Here, the X-ray luminosities are estimated bolometric luminosities for a Universe
with $\Lambda=0.7, \Omega_m=0.3, h=0.7$, assuming gas temperatures of 3 keV.  The $L_X-\sigma$ correlation is consistent
with the local relation, over scales ranging from the richest clusters \citep{D+93,Markevitch}
to groups \citep{XW}.  It is now well known that this scaling is inconsistent with
a purely self-similar model of the intracluster medium, but can be successfully matched
by models with a substantial entropy floor due, perhaps, to the injection of energy from supernovae
and AGN \citep{Babul2}.  However, our uncertainties on $\sigma$ are too large to
improve the existing constraints on the slope of this relation.

\begin{figure}
\leavevmode \epsfysize=8cm \epsfbox{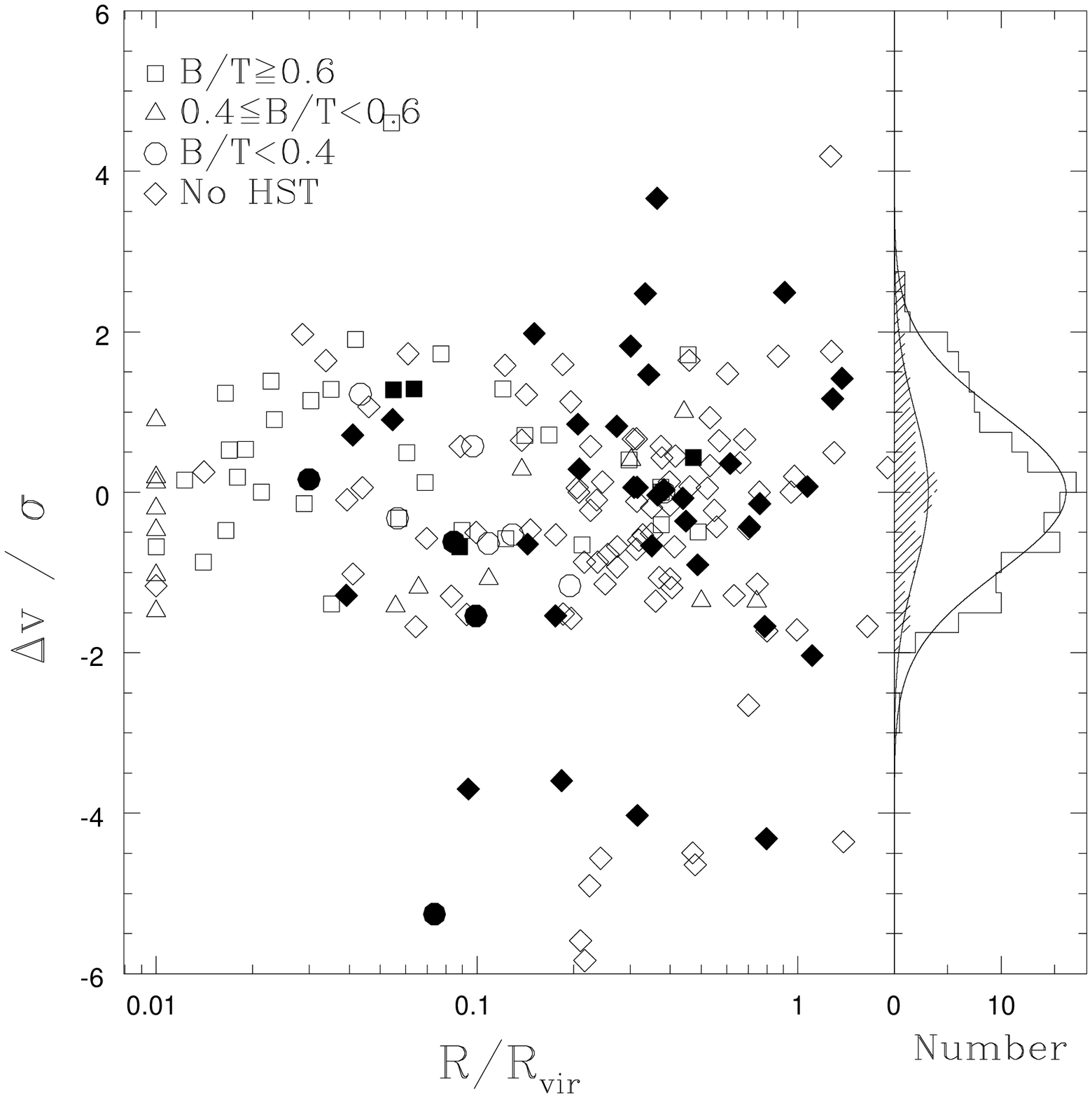}
\caption{The normalised velocity-radius relation for galaxies in the ten clusters.  The velocities
are measured relative to the cluster redshift, normalised to the cluster velocity
dispersion.  The radii are measured from the position of the central, bright galaxy
(shown at $R/R_{\rm vir}=0.01$ for display purposes), and normalised to the cluster virial
radius.  {\it Filled symbols} represent galaxies with \ewoii$>5$\AA.  Only galaxies
within 6 times the cluster velocity dispersion $\sigma$ are shown; cluster members are selected to
be those within 3$\sigma$.  The symbol shape corresponds
to the galaxy morphology, as indicated in the legend.  The normalised velocity distribution
for cluster members is shown in the right-hand panel, for the full sample {\it (open histogram)} 
and the emission line galaxies (\ewoii$>5$\AA, {\it filled
histogram}).  Both are consistent with a Gaussian distribution of unit variance, shown as the smooth,
solid curves.
\label{fig-vr}}
\end{figure}

Figure \ref{fig-vr} shows the normalised velocity-radius correlation for the ten
clusters.  The cluster centres are taken to be the position of the brightest galaxy (\S~\ref{sec-BCG}),
which are always near the X-ray centres from \citet{V+98}.   The radius is the distance 
to this centre, normalised to the
cluster virial radius, while the velocity is the velocity difference from the cluster
mean, normalised to the 1-$\sigma$ velocity dispersion.  The cluster members are well-separated
from the surrounding field.  Emission line galaxies and disk-dominated galaxies both
avoid the central regions of the cluster.  However, there is no measurable difference
between the dynamics of the emission line galaxies and the rest of the sample, as shown by
the comparison of the normalised velocity histograms, in the right panel of Figure~\ref{fig-vr}.  Both velocity
distributions are consistent with a Gaussian distribution of unity variance.  

\subsection{Brightest Cluster Galaxies}\label{sec-BCG}
The central galaxies of each cluster are shown in Figure~\ref{fig-BCG}.  
The central galaxy of Cl0818 lies directly behind
a bright, foreground spiral galaxy.  The spectrum shows features from both galaxies, but
is dominated by the foreground galaxy so is omitted from the rest of the analysis.
All the other central galaxies are
giant elliptical galaxies, none of which show emission lines, nor the prominent Balmer absorption
lines that would indicate the presence of recent star formation activity.  Star formation
in central galaxies \citep[e.g. see ][]{C+99} is likely to be associated with cooling flow activity,
which is now known to produce a reservoir of cold molecular gas \citep{Edge_CO}.
Because only extended sources are included in the catalogue of \citet{V+98},
at the faint flux limit the catalogue is biased against clusters with strong cooling flows
(if they even exist at these low luminosities), which may therefore be related to the 
lack of emission in the central galaxies.

\subsection{Spectral properties}\label{sec-em}
\begin{figure}
\leavevmode \epsfysize=8cm\epsfbox{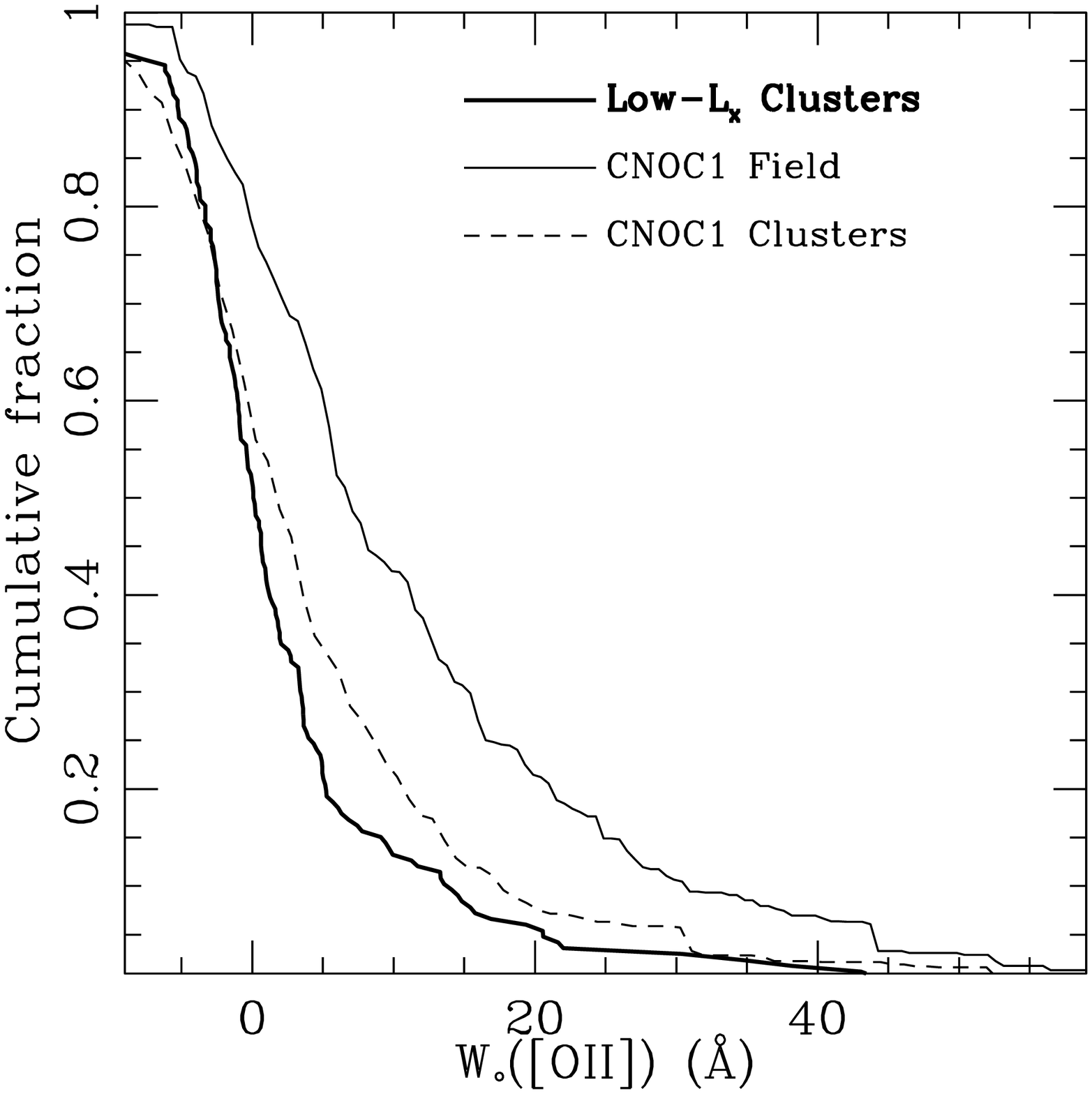}
\caption{The cumulative distribution of \ewoii, for the 167 cluster members
with reliable measurements, brighter than $M_R\sim -18.5+5\log{h}$.  The distribution is weighted by the spectroscopic selection
function.  This is compared with the distribution in
the field and in clusters of high X-ray luminosity at $z\sim 0.3$, from \citet{B+97}.
Both our survey and that of Balogh et al. sample the clusters out to the
virial radius, and are statistically complete at this luminosity.
\label{fig-cnoc}}
\end{figure}

The cumulative distribution of \ewoii\ for the 167 cluster galaxies,
weighted by the spectroscopic selection function, 
is shown in Figure~\ref{fig-cnoc}.  This sample is shown limited to
$M_r\leq -18.5+5\log{h}$, corresponding to $R\sim 20$ at $z=0.25$, to allow
a fair comparison with the field and cluster samples of \citet[][see \S~\ref{sec-discuss}]{B+97}.
The mean (weighted by the selection function) is \ewoii$=3.2$\AA, and the median is \ewoii$=0.7$\AA.
A total of 36 galaxies have \ewoii$>5$\AA; accounting for the spectroscopic selection function,
this corresponds to $\sim 22\%$.  
None of the galaxies have \ewoii$>45$\AA, or otherwise have spectra characteristic
of a strong starburst.  

\begin{figure}
\leavevmode \epsfysize=8cm\epsfbox{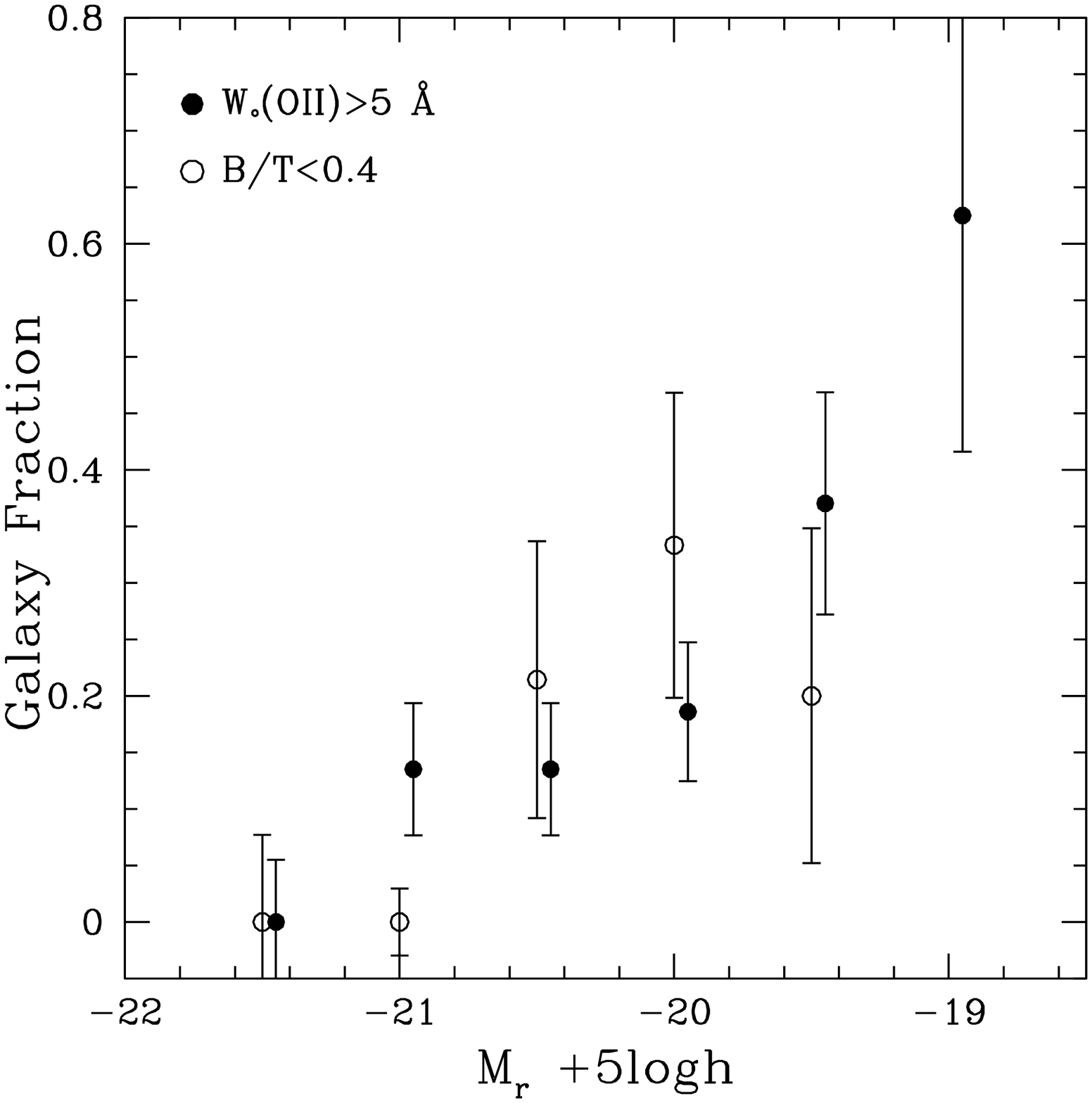}
\caption{The fraction of emission-line galaxies (\ewoii$>5$\AA, {\it solid circles})
and disk-dominated galaxies ($B/T<0.4$, {\it open circles}) as a function of galaxy
luminosity.  Only bins with at least three galaxies are shown.
Error bars are jackknife estimates or, in the case where the fraction
is zero, estimates assuming Poisson statistics.
\label{fig-oiilum}}
\end{figure}

The fraction of emission-line galaxies, therefore, is comparable to the fraction
of disk-dominated galaxies at our spectroscopic magnitude limit (see \S~\ref{sec-morph}).
In Figure~\ref{fig-oiilum} we show these fractions as a function of
luminosity.  The emission line fraction increases strongly with decreasing luminosity,
which is a well-known result \citep[e.g.][]{autofib_lf,L+96,L+99,Christ,IRLF}.  This trend is also seen, with less
significance, in the fraction of disk-dominated galaxies, though the
{\it HST} sample is smaller and morphologies are not measured for the faintest galaxies.
The two fractions are comparable at all luminosities.  However, the 
excess of disk galaxies relative to emission line galaxies around $L^\ast$,
although statistically insignificant, corresponds
to a bona-fide population of ``anemic'' disk galaxies, which we discuss in \S~\ref{sec-anemic}.

\begin{figure}
\leavevmode \epsfysize=8cm \epsfbox{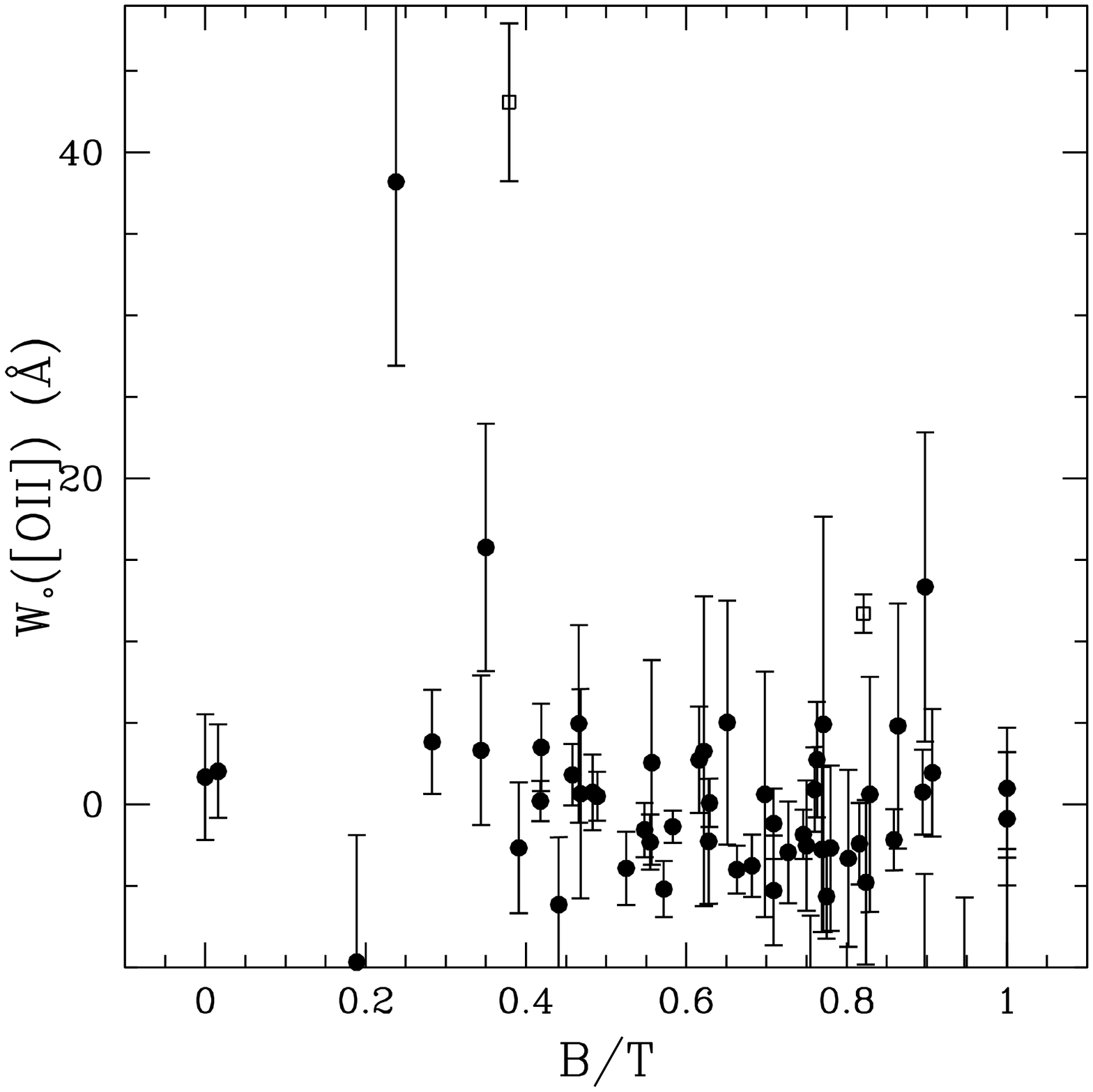}
\caption{The equivalent width of \oii\ as a function of fractional bulge
luminosity, B/T.  The two galaxies represented by {\it open squares} 
have strong, broad [N{\sc ii}] emission characteristic of non-thermal
emission.  
\label{fig-morph}}
\end{figure}

In Figure~\ref{fig-morph} we show the dependence of emission line equivalent width
on B/T.  Only two bulge-dominated galaxies ($B/T>0.6$) show \ewoii$>5$\AA.  One of
these (Cl0841\#38) is a double-nucleated galaxy with strong, broad [N{\sc ii}] indicative
of an active galactic nucleus (AGN).  The other, Cl0849\#20\ is an Sa galaxy with a clear disk component,
close to another bright galaxy which complicates the surface-brightness fitting procedure.
Similarly, of the eight disk-dominated galaxies with $B/T<0.4$, only
three have \ewoii$>5$\AA, and one of these has broad H$\alpha$ and strong [N{\sc ii}]
emission indicative of an AGN.
This fraction of emission-line disk galaxies is consistent with the values
seen in the inner regions ($<0.1 R_{\rm vir}$)
of the high X-ray luminosity CNOC1 clusters \citep{B+98}, and Abell 1689 at
$z=0.18$ \citep{A1689}.   All of these galaxies have luminosities within $\sim0.5$ mag of
$M_r^\ast\sim-20.3+5\log{h}$.

\begin{figure}
\leavevmode \epsfysize=8cm\epsfbox{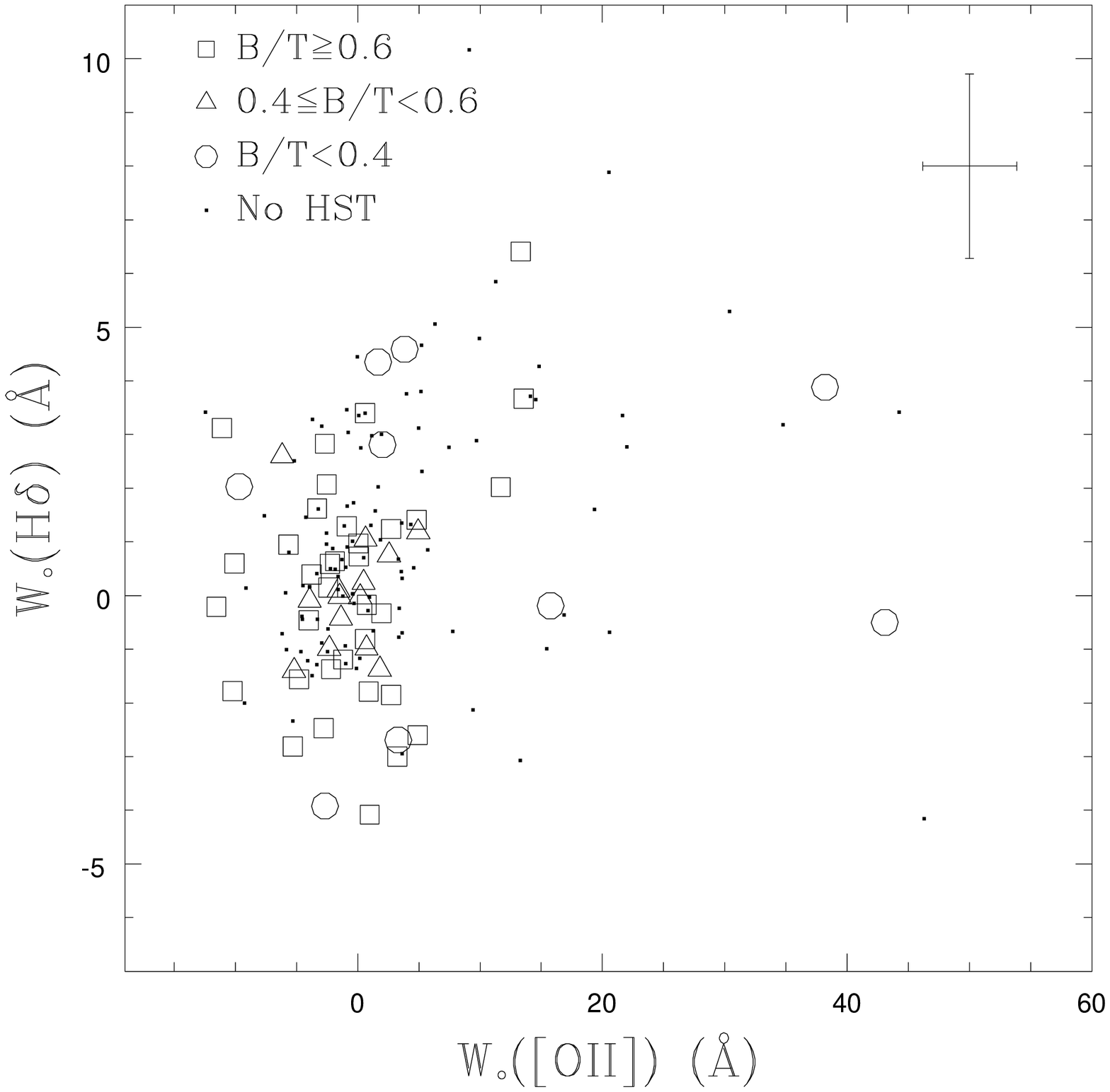}
\caption{The rest frame equivalent width of H$\delta$ is shown as a function
of \ewoii, for all cluster members in which both lines could be measured.
For galaxies with {\it HST} imaging, the symbols correspond to the B/T ratio,
as indicated in the legend.  The sample error bars show the median 1$-\sigma$
uncertainty in each index.
\label{fig-oiihd}}
\end{figure}

While the \oii\ line is present where star-formation is ongoing, the
\hd\ absorption line is expected
to be strong in galaxies in which star formation has occurred sometime within
the last $\sim 500$ Myr or so \citep[e.g.][]{CS87,P+99,PSG,PW00}.  
Measurements of \ewhd\ are shown as a function of \ewoii\ in Figure~\ref{fig-oiihd}.
A population of galaxies with strong H$\delta$ but no detectable
emission are strikingly absent from this sample.
Using the spectral classifications of \citet{D+99}, a k+a galaxy
is one with \ewhd$>3$\AA\ and no detectable emission.  
This definition is somewhat arbitrary, and any physical interpretation of the H$\delta$
strength needs to account for differences in the way in which the line is
measured.   Although fourteen of the
galaxies in our sample formally satisfy Dressler et al.'s definition of a k+a galaxy, 
most of these have \ewhd\ very close to the limit of
3\AA\ (Figure \ref{fig-oiihd}). 
Given the large uncertainties (systematic and random)
in the measurements and the model-sensitivity of the interpretation, we cannot
claim that the spectral properties of this population are strikingly unusual (see \S~\ref{sec-anemic}
for more discussion).
We are only able
to identify four galaxies which show \ewhd$>4$\AA\ with at least
1$\sigma$ confidence.  All of these galaxies show nebular
emission, and thus are e(a) galaxies in the classification scheme of \citet{D+99}.
Both of these populations will be discussed further in \S~\ref{sec-anemic}.

\subsection{Population Gradients}\label{sec-density}
\begin{figure*}
\centerline{\psfig{file=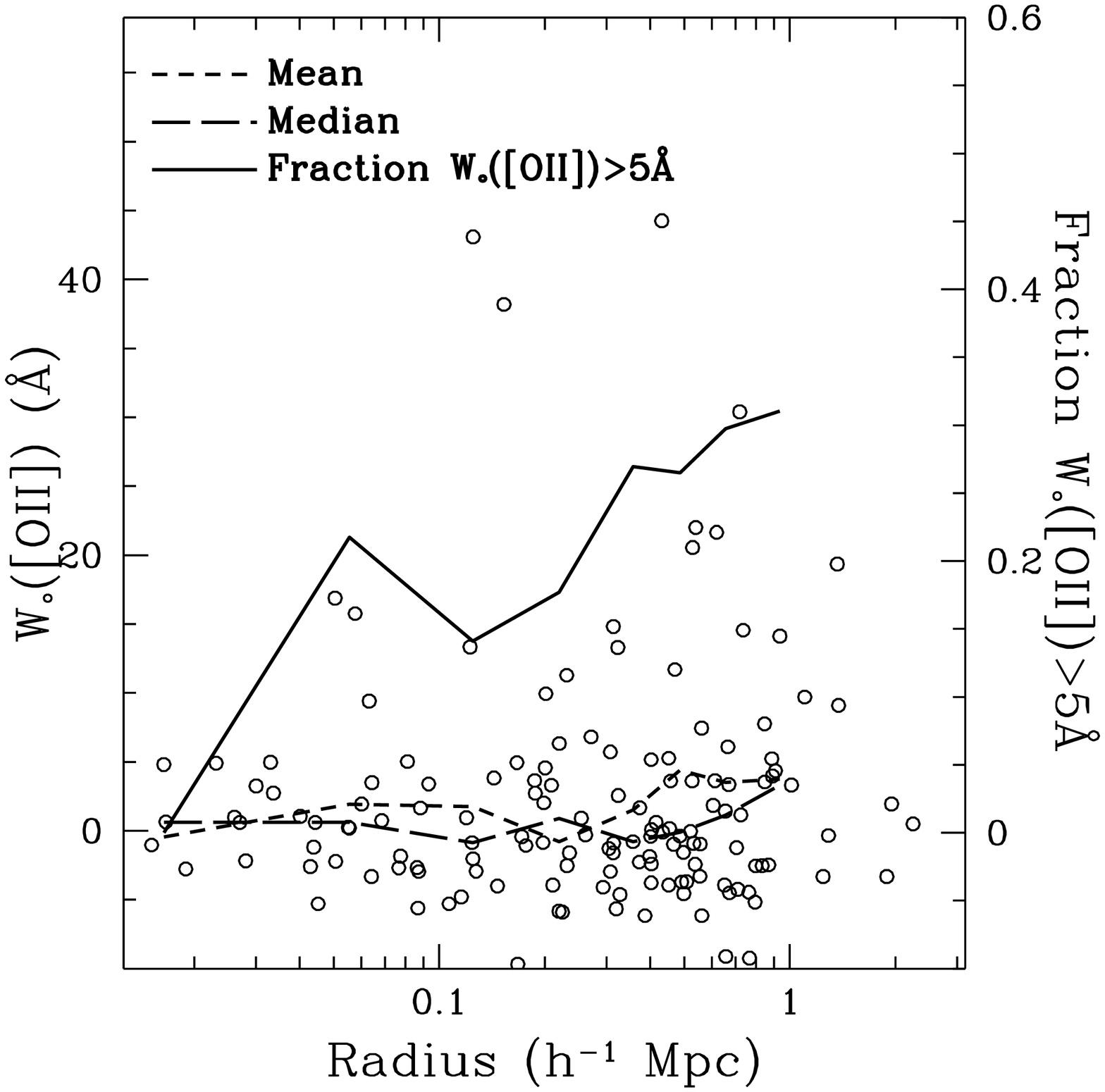,width=3.25in}\hspace*{0.25cm}\psfig{file=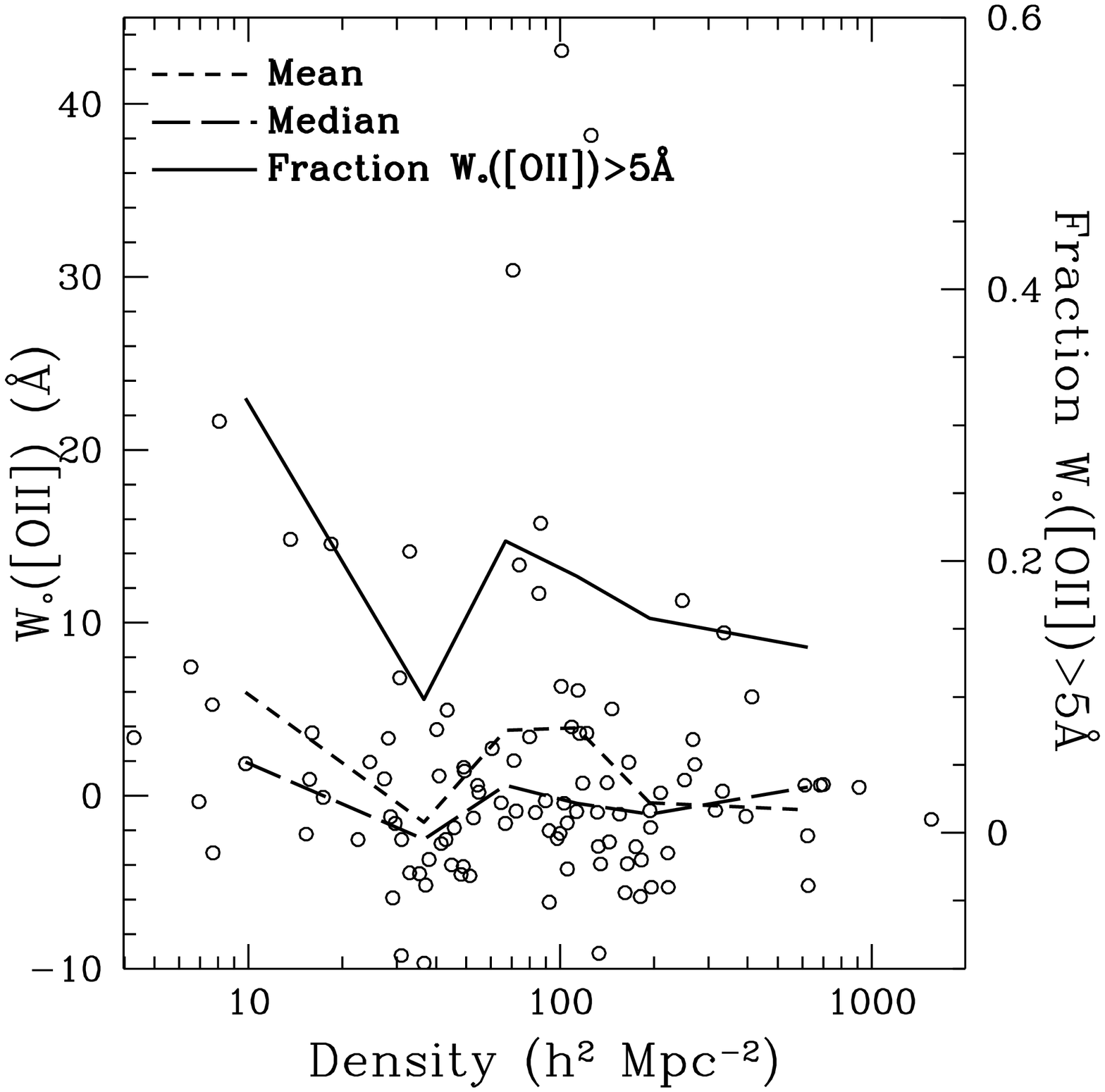,width=3.25in}}
\caption{{\bf Left: }\oii\ emission line strengths as a function of cluster-centric distance.
The {\it long-dashed} and {\it short-dashed} lines show the median and mean value,
respectively, in bins of varying width, each containing 20 points.  The {\it solid} line represents the fraction of galaxies with \ewoii$>5$\AA,
according to the scale on the right side of the figure.  These quantities are
weighted by the spectroscopic selection function, for a sample with $R<20.5$.
{\bf Right: }The same as the left panel, but as a function of local projected density.
The density is computed from the distance to the
fifth nearest-neighbour, and corrected for the background using the number counts of
\citet{CNOC2}.
\label{fig-radius}}
\end{figure*}
In Figure \ref{fig-radius} we show \ewoii\ as a function of cluster-centric radius.
The cluster centres are taken to be the position of the brightest galaxy.
There is little trend in the mean or median equivalent width of the
sample; both are always $\lesssim 5$\AA\ out to $1 h^{-1}$ Mpc.
Galaxies with strong emission lines are only
found outside the cores of these clusters, beyond $\sim 0.05 h^{-1}$ Mpc,
and the fraction of galaxies with \ewoii$>5$\AA\ increases from $<10$\% within
this radius to $\sim 30$\% at 1$h^{-1}$ Mpc, approximately the virial
radius estimated for these systems.  However, this may just be due to the
fact that there are fewer galaxies in the core so the wings of the highly
skewed distribution are insufficiently sampled.  A Kolmogorov-Smirnov test
cannot reject the hypothesis that the \ewoii\ distributions within and
beyond $0.1 h^{-1}$ Mpc are drawn from the same population.

Galaxy morphologies are known to correlate strongly with the local
density of galaxies \citep{Dressler,PG84,DML}.  More recently, a similar
density-dependence has been found for the average star formation rate of galaxies
\citep{C+01,Kodama_cl0939,kap2,2dF-sfr,Sloan_sfr}.  
To compute the local, projected galaxy density, we measure the area enclosing the fifth nearest neighbour to each cluster
member.  We do this using the full photometric catalogue (i.e., not just galaxies
with redshifts) and statistically correct for the average background density.
This is similar to the definition of \citet{Dressler}, but we use the fifth nearest
neighbour rather than the tenth because we found the latter method tends to wash out
the densest regions of the clusters.  Our results are unchanged if we adopt the
tenth nearest-neighbour definition, but the resolution of the dense cores is poorer.
It is important to compute the density to a fixed luminosity limit in all
clusters, and we adopt $M^\ast+1.5$, which corresponds to
$M_r\sim -19.5+5\log{h}$, or $R=20$ at $z=0.25$.  This limit is consistent with that of \citet{Dressler},  
and is 3 mag brighter than that used in Paper~I.
We do not include
the double cluster Cl\,1444 in this analysis, because the projected surface density
cannot be reliably determined from the photometric properties alone.
For the background correction, we use the field number counts in $R_c$
of \citet{CNOC2}.  The number of galaxies brighter than $R_c=20$ 
is 1460$\pm40$ per square degree, where the error does not include cosmic variance.  
This is adjusted as necessary for the magnitude limit
of each cluster.    Finally, we invert the
area containing the fifth nearest neighbour to obtain the density in units of
galaxies per Mpc$^2$.

In the right panel of Figure~\ref{fig-radius} we show the \ewoii\ as a function of local projected
galaxy density.  In the very densest regions, $\Sigma>500$h$^{2}$Mpc$^{-2}$,
there are no galaxies with strong emission lines.  However, below this limit there
is no evidence for a trend with density.  The fraction of galaxies with \ewoii$>5$\AA\ 
remains $\lesssim 20$\%, and the median is $<5$\AA.  

The lack of a correlation of \ewoii\ with density is surprising, especially given the
local result of \citet{2dF-sfr}, that the correlation holds in all clusters,
independent of mass.    This is possibly a consequence of our small sample size, since there are very few galaxies
with emission lines in the full sample.  In particular, in
the lowest density regions we have very few galaxies (13 with $\Sigma<20$),
and thus cannot precisely determine the fraction of emission line galaxies, especially
when that fraction is $\ll1$.

\section{Discussion}\label{sec-discuss}
\subsection{Anemic spirals, starburst and post-starburst galaxies}\label{sec-anemic}
Despite the similarity in the fraction of disk-dominated and emission-line galaxies ($\sim 20$\%),
there is a significant population of disk-dominated galaxies in our sample that
do not have detected emission lines.  This may be analogous to the population seen
in more massive clusters \citep{P+99}, and  is a very cluster-specific population;
in local field samples, almost all disk galaxies show strong emission
lines \citep{K83,Jansen}.   In Figure~\ref{fig-morph} we showed that there
were six galaxies in our sample with
$B/T<0.4$ and \ewoii$<5$\AA.
In one of these galaxies, Cl\,1701\#149, there is H$\alpha$ emission,
\ewha$=23\pm 3$\AA, which is weak enough to be consistent with the
low observed \ewoii.  The remaining five galaxies
are all convincing disk galaxies with no detectable emission.
If we relax our definition of a disk-dominated galaxy to include those with
larger $B/T$ ratios ($B/T<0.5$), we find five other clear examples of galaxies
with a disklike morphology but no evidence of nebular emission.  We also
include a galaxy (Cl\,1309\#119) without \ha\ emission (\oii\ is undetermined)
to bring the total sample of such galaxies to eleven.  This population therefore comprises
$6.5 \pm 2$\% of the cluster members, and $57 \pm 17$ \% of the disk-dominated ($B/T<0.5$) population.
These are consistent with the fractions of late-type (later than Sa) galaxies without
emission found in the z$\sim 0.4$ cluster sample of \citet{P+99}.
The images and spectra of these galaxies in our clusters are shown in Figure~\ref{fig-anemic}.
Most appear to be early or mid-type spirals, but 
have very smooth disks, without strong spiral structure
or prominent H{\sc ii} regions, and thus resemble anemic galaxies \citep{vdB76,vdB91}.
Particularly interesting is Cl0818\#58, a smooth spiral
galaxy that is elongated and asymmetric, and has the appearance expected of a galaxy
in the process of being stripped \citep{QMB}.  We note that the slit width is $1.5\arcsec$, half the
width of the postage-stamp images shown in Figure~\ref{fig-anemic} and comparable to the size of the
galaxies; therefore we do not expect the lack of emission to be due to an aperture bias. 

\begin{figure}
\leavevmode \epsfysize=8cm\epsfbox{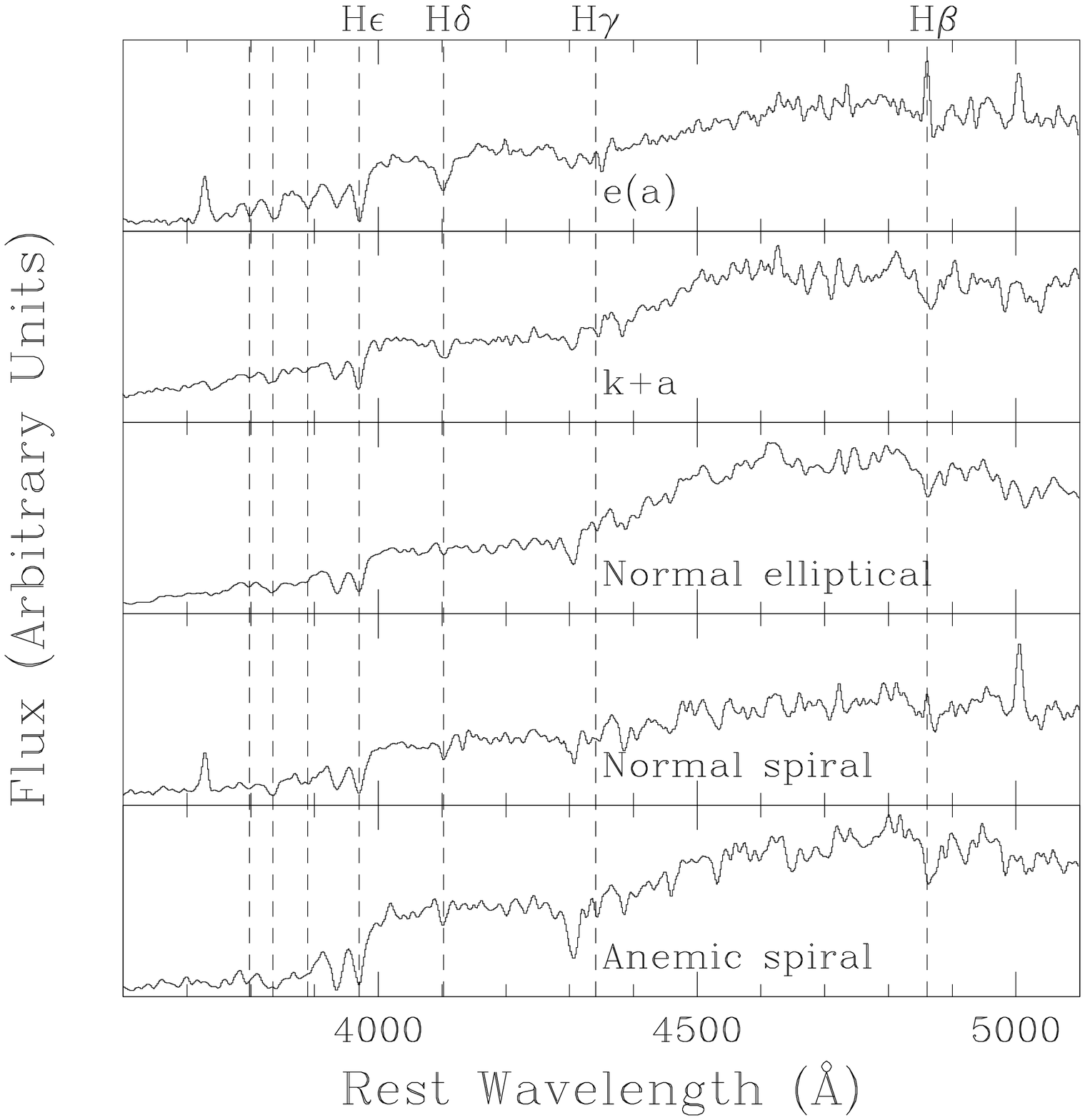}
\caption{Coadded spectra of 11 anemic spiral
galaxies ($B/T<0.5$ and no emission lines), 13 normal spiral galaxies,
34 elliptical galaxies ($B/T>0.6$), 14 k+a galaxies (\ewhd$>3$\AA\ and
\ewoii$<5$\AA), and 8 e(a) galaxies.  The spectra
are renormalised to template continuua from \citet{Kinney} as described in the text,
and smoothed to the instrumental resolution of $\sim 15$\AA.   The positions of the Balmer
absorption lines are marked with {\it dashed}, vertical lines.
\label{fig-coadd}}
\end{figure}

To improve upon the signal-to-noise ratio of individual spectra, we have coadded four
classes of spectra, in Figure~\ref{fig-coadd}.  In particular, we show the coadded spectra
of the anemic spiral sample, including all galaxies with $B/T<0.5$ and no detectable emission
lines.  Before coadding, the individual spectra are shifted to zero
redshift and the shape of the continuum in the range $3500$--$5100$\AA\ is removed with
a spline fit.  The
spectra are then averaged pixel by pixel, weighted by the median flux in
the rest-frame wavelength range $4050$--$4250$\AA, to give more weight to better quality data.
For presentation purposes, we fit the continuum of the Sb template spectrum
from \citet{Kinney}, and rescale our spectra to this continuum.  Finally, the spectra are
smoothed to the instrumental resolution of $\sim 15$ \AA.

The coadded anemic spiral spectrum can be directly compared with the coadded spectra of
thirteen normal spiral galaxies (disk galaxies with emission lines, including clear examples of spiral
galaxies in the absence of {\it HST} imaging) and 34 early-type galaxies (with {\it HST} imaging and $B/T>0.6$), also 
shown in Figure~\ref{fig-coadd}.   The early-type galaxies are renormalised to the continuum
of the elliptical template of \citet{Kinney}.  Apart from the lack of emission lines, the
anemic spiral spectra look quite similar to those of the normal spiral galaxies.  In particular,
the Balmer series is stronger than seen in the early-type spectra, comparable to that of normal
spirals (\ewhd$\sim 2.5$\AA).  On the other hand, some absorption lines in the anemic galaxies, most notably the G-band at $\sim 4300$\AA,
are more similar to the strengths seen in elliptical galaxies.

It is remarkable that none of the galaxies in our sample show the very strong emission
lines characteristic of a strong starburst.  The strongest emission-line galaxies are almost
all normal spiral galaxies (with the exception of the merging/interacting galaxies Cl0841\#38
and Cl0849\#20), and the fraction of strong emission-line galaxies is small relative to the fraction
observed in the field at this redshift.  Furthermore, as shown in \S~\ref{sec-em}, we find no convincing examples of
the Balmer-strong galaxies without emission lines which may be in a post-starburst phase
\citep[e.g.][]{CS87,DG83,PSG,P+99}.  Formally, there are fourteen galaxies which are classified k+a
according to \citet{D+99}, though all but three of these have \ewhd\ $\sim 3$\AA, close to the
arbitrarily-defined cutoff strength.  The coadded spectrum of these galaxies is shown in Figure~\ref{fig-coadd}.
No evidence for H$\beta$ or [O{\sc iii}]$\lambda5007$ emission is seen, confirming that the lack of \oii\
emission is real.  Most of the spectrum appears similar to that of the elliptical population although,
by selection, the H$\delta$ absorption line is relatively strong.
Most notably, the other Balmer lines are not especially prominent, which suggests that many of
the high H$\delta$ measurements are cases where the substantial measurement uncertainty results in
an overestimate of the line strength, as suggested by \citet{PSG}.
We do find eight galaxies with \ewhd$>4$\AA, though the errors are such that only
four of these exceed 4 \AA\  with $>1\sigma$ confidence.  All of these galaxies have emission lines, and
are thus e(a) galaxies \citep{P+99}.  The coadded
spectrum of these eight galaxies is also shown in Figure~\ref{fig-coadd}.  The Balmer series
is clear even blueward of H$\epsilon$, and is enhanced relative to that seen in the normal
spiral, anemic spiral, or k+a population.  Emission lines at H$\beta$ and [O{\sc iii}]$\lambda5007$
are seen, in addition to \oii.  

The rarity of galaxies 
with strong Balmer absorption lines is in
good agreement with the results seen in  X-ray luminous clusters
at $z\sim 0.3$, from the work of \citet{PSG}.  
The discrepancy with the relatively
high fraction of H$\delta$-strong galaxies in clusters at $z\sim0.4$ \citep{P+99}
is still not understood.  One suggestion that has been put forward is that the clusters of Poggianti et al.
are dynamically younger, with more galaxy-galaxy interactions than in the more relaxed
CNOC1 clusters \citep{PSG}.  However, we expect
such interactions to be even more important in our sample of clusters, due to their low
velocity dispersions, and yet no large population of starburst or post-starburst galaxies
is found.  The fact that these clusters are X-ray selected and have central, giant elliptical
galaxies, may 
suggest that they are dynamically old systems, in which all merger-induced
star formation activity took place several Gyr ago.  However, we still cannot
rule out the possibility that the difference is due in part to evolution between
$z\sim0.25$ and $z\sim 0.4$ (corresponding to a difference of 1.3 Gyr in our assumed
cosmology), 
or to spectroscopic sample selection effects, as discussed in \citet{PSG}. 

\subsection{Comparison with X-ray luminous clusters}
The ten clusters analysed in this work are analogous to more massive clusters
in several ways.  Dynamically, they appear to be relaxed systems in which the X-ray
centre coincides closely with the position of the giant elliptical galaxy.  Interestingly,
none of these central galaxies show any sign of star formation.  In contrast, in 
the {\it ROSAT} Brightest Cluster sample of clusters, $\sim 27\%$ have central galaxies
which show emission lines, approximately independent of X-ray luminosity \citep{C+99}.
The clusters in our sample have luminosities which place them all in the lowest-luminosity
bin of Crawford et al.'s figure 4, where  $\sim 22\pm10$\% have central galaxies
with H$\alpha$ emission.  Thus, our results for a sample of nine clusters (omitting Cl0818, for
which no spectrum of the central galaxy was obtained) are not strongly
inconsistent with this fraction.

In massive clusters, \citet{CNOC_dynamics} found that the velocity dispersion of blue
galaxies is $\sim 30$\% larger than that of the red galaxies, and suggested this was a sign that the
star-forming galaxies are not yet in virial equilibrium with the cluster potential.  A similar conclusion
was reached by \citet{D+99}, who found that recently star-forming galaxies have a velocity dispersion
$\sim 40$\% larger than that of the passive, elliptical population.
We find no statistically significant difference in the dynamics of the emission-line galaxy population,
relative to that of the whole population.  However, our sample of emission line galaxies is too
small to claim a significant difference from the more massive clusters.  Using a Kolmogorov-Smirnov
test, we can only rule out, at the 99\% or greater confidence level, velocity distributions
(for the emission-line population) that are more than 2.8 times broader than the cluster 
velocity dispersion.

The distribution of \ewoii\ in our cluster sample (Figure~\ref{fig-cnoc}) is
similar to that of \citet{B+97}, measured, to the same luminosity limit, for the CNOC1 sample of X-ray luminous 
clusters at $z\sim 0.3$ \citep{YEC,CNOC1}.  Both our
sample and the high X-ray luminosity clusters show \ewoii\ distributions
that are greatly 
suppressed relative to the field near that redshift, also taken from \citet{B+97}.
Note that, since it
is generally easier to get redshifts for emission line galaxies, any incompleteness in
our sample for this reason is likely
to lead to our overestimation of the number of emission line galaxies in our sample, thus
strengthening our conclusions.
Therefore, we conclude that over $\sim 2$ orders of magnitude in cluster X-ray luminosity, there is little difference
in the mean age of the stellar population for galaxies more luminous than $\sim M^\ast+2.5$.
It must, therefore, be a local process, rather than a global one associated with the
large-scale mass distribution, which affects the star formation rates of galaxies.

Both the present study and the CNOC1 cluster sample present data out to approximately the virial
radius of the clusters.  
However, the CNOC1 clusters are more massive, and
there may be a difference in the range of local densities sampled.  
To test this, we have evaluated the local density for every galaxy in the CNOC1
sample, using the same method, and the same luminosity limit, as for our \lowlx\
cluster sample.  
In Figure~\ref{fig-cnocden} we show the fraction of galaxies with \ewoii$>5$\AA\
as a function of local density, for the present sample and that of \citet{B+97}.
The two functions are indistinguishable within the uncertainties, and thus we conclude that the level of star
formation in the \lowlx\ clusters is comparable to that in the CNOC1 clusters
at all densities probed. In particular, we see little trend with local density,
and the fraction of emission-line galaxies is always $<30$\%.  

In Paper~I
we showed that the \lowlx\ clusters have a significantly larger fraction
of disk-dominated galaxies brighter than $R\sim 23$ than more massive clusters (Paper~I).  
In particular, the data showed that, at a given local density, the bulges in massive
clusters are systematically more luminous than the bulges in the \lowlx\ clusters, while
the disk  luminosity function is independent of cluster mass. 
In the present
work we do not find a large difference between the fraction of emission-line galaxies
at $R<20$.   
This may suggest that galaxy
morphology (in particular, bulge size) is partially sensitive to large-scale structure, 
while star formation properties are not.  Alternatively, this may just be reflecting the difference 
in the luminosity ranges considered
in the two studies; unfortunately, the {\it HST} sample is too small to
determine the disk fraction at $R<20$ with enough precision to determine whether or not
the small difference found in Paper~I holds at this brighter magnitude.

Very recently, analysis of the correlation between star formation rate and local
projected density in nearby clusters from the 2dF galaxy redshift survey \citep{2dF-sfr}
and the Sloan Digital Sky Survey \citep{Sloan_sfr} has shown that star formation
is reduced below the global average in all environments where the local density exceeds
$\sim 1$ galaxy (brighter than $M^\ast+1$) per Mpc$^{2}$.  This was shown to hold
in systems of low velocity dispersion, and well outside the virialised cluster regions.
Although our sample does not extend to such low densities, we confirm that, even at $z\sim 0.25$,
where the Butcher-Oemler effect \citep{BO84,Erica} is beginning to appear, dense regions in
low-mass structures have very low star formation rates\footnote{Deriving a star formation rate from the \ewoii\
requires knowledge of the galaxy $B-$band luminosity, metallicity and dust content.  If these
quantities in our clusters are similar to those in the $z=0$ clusters, than the similarity in
\ewoii\ distributions corresponds to a similarity in star formation rate distributions.}.
It will be of enormous interest
to trace the star-formation rate correlation with density out to comparably low densities
at this redshift and beyond, to compare with the low redshift data.

\subsection{Comparison with the morphology-density relation}
\begin{figure}
\leavevmode \epsfysize=8cm \epsfbox{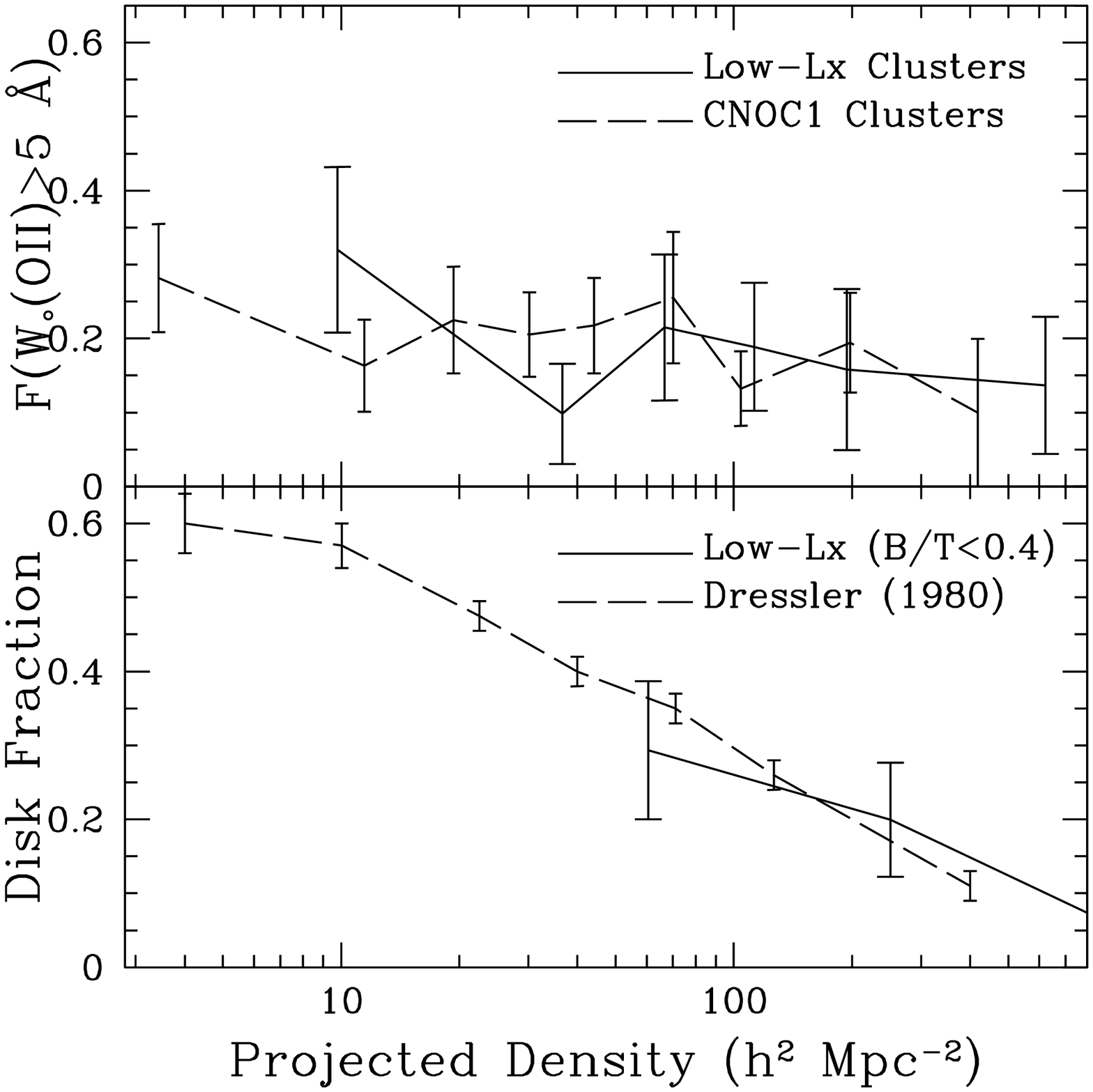}
\caption{
{\bf Top: }The fraction of galaxies with \ewoii$>5$ \AA\ in the
present sample ({\it solid line}) and in the CNOC1 sample \citep{B+97},
as a function of local, projected galaxy density.
Error bars are 1-$\sigma$ jackknife estimates.
The fractions corresponding to the present sample are computed in equally populated
bins containing 20 galaxies.  The CNOC1 data are presented in bins each with 40
galaxies.
{\bf Bottom: } The fraction of disk galaxies ($B/T<0.4$) from our {\it HST} sample
as a function of local density ({\it solid line}) is compared with the fraction of
spiral and irregular galaxies from \citet[][{\it dashed line}]{Dressler}.
\label{fig-cnocden}}
\end{figure}

It is interesting
to compare the \ewoii\ dependence (or lack-thereof) on density  
with the morphology-density relation \citep{Dressler}.  We show the measured disk
fraction for our sample in this region as the heavy, dashed line in Figure~\ref{fig-cnocden}.   
Since our {\it HST} imaging
is restricted to the central regions of the clusters, reliable morphologies are 
only available for the densest regions.  To make a comparison at lower densities, we
show the fraction of spiral and irregular galaxies in the sample of \citet{Dressler}, as a function of
local density, converted to $h=1$ for consistency with the results shown here. 
At the high density end, Dressler's data are consistent
with the disk fractions we measure from the {\it HST} data.

There is a much stronger trend compared with the \ewoii\ data, and the fraction
of spiral galaxies increases to $\sim 60$\% at the
lowest densities probed by our spectroscopic sample. 
At densities $\lesssim 50$ Mpc$^{-2}$, the fraction of
spiral galaxies expected from Dressler's data is at least a factor of two larger than the fraction of
galaxies with \ewoii$>5$\AA, significant at the $>2\sigma$ level.  
In agreement with the results of \citet{B+98}, \citet{C+01} and \citet{2dF-sfr}, this suggests
that the morphology-density relation is at least partially independent of the
star-formation rate dependence on density.   

\subsection{Physical mechanisms}
We can use these results to draw some conclusions about the physical
mechanisms which may be responsible for the low star formation rates among
galaxies in dense environments.
The clusters in this sample have velocity dispersions which are typically 
a factor $\sim 2$ less than those of the most massive clusters in the universe at
$z\sim 0.25$.  Since the ram-pressure force on a galaxy travelling through the
intracluster medium is proportional to $v^2$ \citep{GG}, we expect ram-pressure stripping of
disk gas to be $\sim 4$ times less effective in our cluster sample than in more
massive clusters.  In particular, models suggest that ram-pressure effects should be
negligible in clusters with virial temperatures $kT\sim 2$ keV \citep{FN99}, corresponding
approximately to the expected temperatures of our \lowlx\ cluster sample. 
However, this difference is not reflected in the current star
formation rates of the observed galaxy population.  Therefore, we conclude that ram-pressure
stripping within clusters is not primarily responsible for the low star formation
rates.  This is in good agreement with the results of other studies, which have shown
little or no trend in blue galaxy fraction with cluster X-ray luminosity
\citep{F+01,kap2}.

Secondly, we have not found a population of starburst galaxies, nor evidence
for a large population of galaxies which have had a recent burst.  This is in
agreement with more complete H$\alpha$ studies of more massive clusters at $z\sim 0.3$
\citep{C+01,A1689}.  Thus, starbursts induced by the cluster environment (whether by
interactions with the intracluster gas or with other galaxies) do not appear to be
an important process, at least at the epoch at which the clusters are observed.  
There remain two appealing scenarios which are consistent
with our data.  The population gradients observed
in rich clusters are consistent with models of ``strangulation'', in which satellite galaxies in haloes of any
mass are stripped of their hot gas and consequently consume their available fuel supply
fairly gradually \citep{LTC,infall,Diaferio,Okamoto,Bekki02}.  
The appearance of the disk-dominated galaxies which show no sign of star formation (Figure \ref{fig-anemic})
seems to support this hypothesis.  With the possible exceptions of Cl0818 \#58, which has
a highly asymmetric appearance one might expect of a galaxy interacting strongly with the ICM,
and Cl\,1309\#119, which may be interacting with a larger galaxy, all of these galaxies are
fairly isolated, spiral galaxies with smooth disks and no sign of strong disturbance or
bright H{\sc ii} regions.
The other possibility is that the galaxy transformation occurs in even smaller systems ---
galaxy groups --- in the cluster infall region \citep{ZM98,Kodama_cl0939}.  In particular,
at higher redshift galaxy groups are denser systems, with larger velocity dispersions than
their local counterparts, and even processes like ram-pressure stripping may be able to
take place \citep{Fujita-rps}.  Furthermore, this may be the environment in which
``strangulation'' itself is most effective, and subsequent evolution within clusters
may make little difference to the observable properties of the population \citep{Okamoto}.
The next step is therefore to focus on galaxy groups at a series of redshifts; in particular,
galaxies in environments close to the ``critical density'', at which environmental
effects first become observable \citep{Kodama_cl0939,2dF-sfr,Sloan_sfr}.


\section{Conclusions}\label{sec-conc}
We have analysed ground-based spectroscopy and {\it HST}-based morphologies of galaxies
in ten clusters at $z\approx0.25$ with low X-ray luminosities.  The sample includes
165 galaxies brighter than $R=20.5$, which corresponds to $M_r=-19+5\log{h}$
at $z=0.25$.  We have measured morphologies using the {\sc gim2d} surface-brightness fitting
software of \citet{Gim2d}, and the strengths of important spectral features, in particular the \oii\ emission
line.  The properties of the ten clusters can be summarized as:

\begin{itemize}
\item All ten clusters host a giant elliptical galaxy near the centre of the X-ray emission.
None of the nine central galaxies for which we have a reliable spectrum show emission lines.
\item Apart from the double cluster Cl\,1444, and the close pair of clusters Cl\,1701 and Cl\,1702,
all cluster velocity dispersions are consistent with a single Gaussian, and they
appear dynamically well-separated from the surrounding field.  Thus, they appear to be
evolved systems in approximate dynamical equilibrium.
\item The measured velocity dispersions range from $\sim 350$--$850$~km~s$^{-1}$, and
are consistent with the local $L_X-\sigma$ relation observed in larger samples
\citep[e.g.][]{XW}.
\item The fraction of cluster galaxies with \ewoii\ $>5$\AA\ (2$\sigma$ confidence limit)
is 22$\pm 4$\%.  The mean is \ewoii$=3.2$\AA, and the median is \ewoii$=0.7$\AA.
There is no evidence for a significant correlation between \ewoii\
and either radius or density, apart from the lack of strong emission-line galaxies
in the densest, central regions ($\lesssim 0.1$ h$^{-1}$ Mpc).  Also, we do not measure
a significant difference in the dynamics of the emission-line
galaxies, relative to the rest of the population.
\item Disk-dominated galaxies ($B/T<0.4$) comprise 18$\pm5$\% of the sample within the
central 0.4 h$^{-1}$ Mpc covered by our {\it WFPC2} images.  Less than 25\% (2/8)
of these galaxies show significant emission.  The remainder, a population of ``anemic'' disk galaxies, are relatively
isolated, regular spiral galaxies near $L^\ast$, with smooth disks.  Such galaxies
are rarely found in local, field samples, but are also seen, in similar abundance, in
more massive clusters \citep{P+99}.  
\item No galaxies in our sample have a spectrum characteristic of a post-starburst or
of truncated star formation.  Only four galaxies have \hd$>4$\AA\ with at least 1$\sigma$
significance, and all of these show nebular emission lines.  Thus there is no evidence
that these cluster environments act to enhance star formation activity, even temporarily.
\end{itemize}

The distribution of \ewoii\ in these clusters is similar to that measured in
the sample of \citet{B+97}, which is comprised of clusters approximately an order of
magnitude more massive.  Galaxies in both systems show low star formation rates
even at projected surface densities as low as $\sim 10 h^{2}$ Mpc$^{-2}$, where the
fraction of spiral and irregular galaxies expected from the morphology-density
relation is $\sim 60$\%.  The fact that star formation rates are so low even in these
low-mass structures has important implications for understanding galaxy evolution in
general. The phenomenon is not likely to be driven by extreme processes, such
as ram-pressure stripping, or interaction-induced starbursts, which are expected to be important only
in the richest clusters.  Rather it is something that operates in more commonplace 
environments, possibly groups in the infall regions of clusters \citep{Kodama_cl0939,2dF-sfr}.
Tracing the evolution of galaxies in these groups, therefore, may shed light on the
processes responsible for the observed decline in the globally-averaged star formation
rate of the Universe.  


\section*{Acknowledgements}
We thank the referee, Chris Collins, for his expeditious report and 
useful suggestions which improved this paper.  We are also grateful
to the CNOC1 collaboration for allowing us to use their unpublished data.
We acknowledge financial support from PPARC (MLB, RGB, RLD), the Royal
Society (IRS), Leverhulme Trust (IRS, RLD), the Deutsche
Forschungsgemeinschaft and the Volkswagen foundation (BLZ,AF). 
The data in this paper includes observations made with the NASA/ESA {\it Hubble Space
Telescope} obtained at the Space Telescope Science Institute, which is
operated by the Association of Universities for Research in Astronomy
Inc., under NASA contract NAS 5-26555.
\bibliography{ms}
%
%
{\scriptsize
\begin{table*} 
\begin{center} 
\caption{\centerline {\sc \label{tab-data} Properties of Cluster Members}} 
\vspace{0.1cm}
\begin{tabular}{lcccccccrrr} 
\hline\hline
\noalign{\smallskip}
Cluster&  Galaxy ID & $R_{\rm mag}$&Redshift & R.A.             & Dec          & B/T     & S/N  &\ewoii & \ewhd & \ewha \cr
       &            &              &         &\multispan{2}{\hfil (J2000)\hfil}&         &      &{\AA\hfil}   & {\AA\hfil}   &{\AA\hfil} \cr 
\noalign{\smallskip}
\hline
Cl0818 & 13 & 20.5 & 0.2682 &   08 19 17.822   &   56 53 17.833   & ... & 9.6 & $34.7\pm$6.1 & $3.2\pm$2.5 & $80.1\pm$24.4 \cr
Cl0818 & 51 & 20.1 & 0.2723 &   08 19 19.288   &   56 53 56.929   & ... & 6.4 & $20.5\pm$7.4 & $7.9\pm$2.4 & ... \cr
Cl0818 & 54 & 19.6 & 0.2670 &   08 18 56.598   &   56 54 10.346   & $0.02$ & 16.4 & $2.0\pm$2.9 & $2.8\pm$1.5 & ... \cr
Cl0818 & 58 & 20.1 & 0.2667 &   08 18 52.914   &   56 54 25.988   & $0.19$ & 7 & $-9.7\pm$7.8 & $2.0\pm$3.6 & $22.1\pm$11.6 \cr
Cl0818 & 64 & 19.0 & 0.2703 &   08 19 02.249   &   56 54 28.199   & ... & 8.6 & $9.4\pm$5.5 & $-2.1\pm$2.1 & ... \cr
Cl0818 & 66 & 19.6 & 0.2698 &   08 18 54.792   &   56 54 44.158   & $1.00$ & 11.8 & $1.0\pm$3.7 & $-4.1\pm$2.3 & ... \cr
Cl0818 & 67 & 19.0 & 0.2688 &   08 19 09.943   &   56 54 44.773   & $0.77$ & 8.8 & $-2.8\pm$5.1 & $-2.5\pm$3.0 & ... \cr
Cl0818 & 86 & 19.7 & 0.2643 &   08 19 14.008   &   56 55 24.395   & ... & 4.6 & $6.8\pm$22.5 & ... & ... \cr
\smallskip
Cl0818 & 87 & 19.8 & 0.2643 &   08 18 59.467   &   56 55 29.078   & $0.24$ & 5.4 & $38.2\pm$11.3 & $3.9\pm$3.0 & $99.9\pm$10.0 \cr
Cl0819 & 2 & 18.5 & 0.2278 &   08 18 48.715   &   70 52 55.844   & ... & 22.6 & $-3.3\pm$1.5 & $-0.4\pm$0.8 & $3.1\pm$1.1 \cr
Cl0819 & 4 & 19.8 & 0.2290 &   08 20 29.912   &   70 53 06.119   & ... & 11.4 & $21.6\pm$4.0 & $3.3\pm$1.4 & $37.9\pm$3.4 \cr
Cl0819 & 6 & 18.5 & 0.2319 &   08 19 06.486   &   70 53 04.114   & ... & 23.8 & $-0.4\pm$1.5 & $1.0\pm$0.7 & $0.8\pm$0.9 \cr
Cl0819 & 9 & 18.7 & 0.2280 &   08 20 13.008   &   70 53 34.627   & ... & 14.2 & $1.4\pm$3.0 & $1.6\pm$1.3 & $-1.3\pm$1.3 \cr
Cl0819 & 11 & 18.7 & 0.2304 &   08 19 07.756   &   70 53 37.702   & ... & 18.8 & $-4.6\pm$1.8 & $-1.0\pm$1.0 & $-0.6\pm$1.0 \cr
Cl0819 & 12 & 18.7 & 0.2309 &   08 19 37.969   &   70 53 49.513   & ... & 16.4 & $-1\pm$2.5 & $-1.3\pm$1.1 & $0.6\pm$1.1 \cr
Cl0819 & 13 & 18.4 & 0.2294 &   08 19 46.923   &   70 53 53.855   & ... & 20 & $6.1\pm$2.5 & ... & ... \cr
Cl0819 & 15 & 18.7 & 0.2305 &   08 19 35.251   &   70 54 02.423   & ... & 19 & $-4.6\pm$2.1 & $-0.4\pm$1.0 & $1.8\pm$1.0 \cr
Cl0819 & 16 & 18.4 & 0.2320 &   08 19 04.374   &   70 53 54.953   & $0.75$ & 24.2 & $-1.8\pm$1.5 & $0.6\pm$0.7 & $5.1\pm$0.9 \cr
Cl0819 & 20 & 19.1 & 0.2305 &   08 20 12.887   &   70 54 29.862   & ... & 14.6 & $1.9\pm$2.6 & $3.0\pm$1.2 & $11.2\pm$1.7 \cr
Cl0819 & 21 & 18.3 & 0.2272 &   08 19 42.125   &   70 54 32.087   & ... & 25.4 & $-2.5\pm$1.5 & $-1.0\pm$0.7 & $-0.1\pm$0.8 \cr
Cl0819 & 25 & 19.2 & 0.2296 &   08 18 56.120   &   70 54 39.445   & ... & 13.4 & $-2.5\pm$2.8 & $1.2\pm$1.3 & ... \cr
Cl0819 & 26 & 19.7 & 0.2280 &   08 19 22.859   &   70 54 44.914   & ... & 5.4 & $-5.8\pm$6.4 & $-1.0\pm$4.5 & ... \cr
Cl0819 & 27 & 19.0 & 0.2277 &   08 19 01.384   &   70 54 51.642   & $0.52$ & 15.4 & $-3.9\pm$2.3 & $-0.1\pm$1.2 & $-1.6\pm$1.5 \cr
Cl0819 & 29 & 19.2 & 0.2297 &   08 18 54.033   &   70 54 52.628   & ... & 18 & $14.1\pm$2.2 & $3.7\pm$0.9 & $29.8\pm$2.2 \cr
Cl0819 & 33 & 18.6 & 0.2289 &   08 19 17.747   &   70 55 00.952   & ... & 16.6 & $-5.6\pm$2.5 & $0.8\pm$1.1 & $-0.2\pm$1.0 \cr
Cl0819 & 34$^{a}$ & 17.9 & 0.2304 &   08 19 18.340   &   70 55 04.411   & ... & 34.8 & $-1.8\pm$1.0 & $0.5\pm$0.5 & $1.1\pm$0.6 \cr
Cl0819 & 36 & 19.4 & 0.2304 &   08 19 23.751   &   70 55 24.272   & ... & 11.6 & $-0.9\pm$3.7 & $1.7\pm$1.6 & $-3.3\pm$1.6 \cr
Cl0819 & 37 & 19.0 & 0.2310 &   08 19 05.949   &   70 55 34.460   & $0.44$ & 9.2 & $-6.2\pm$4.1 & $2.6\pm$2.5 & ... \cr
Cl0819 & 40 & 18.4 & 0.2274 &   08 19 11.986   &   70 55 43.828   & ... & 23.4 & $-0.4\pm$1.6 & $0.0\pm$0.8 & $0.8\pm$0.9 \cr
Cl0819 & 43 & 19.1 & 0.2313 &   08 19 12.638   &   70 56 05.798   & ... & 14.4 & $-2\pm$2.7 & $0.9\pm$1.3 & $0.7\pm$1.3 \cr
Cl0819 & 44 & 19.9 & 0.2301 &   08 20 09.086   &   70 56 17.527   & ... & 10.6 & $22.0\pm$4.3 & $2.8\pm$1.6 & $64.9\pm$6.0 \cr
\smallskip
Cl0819 & 47 & 18.2 & 0.2306 &   08 19 23.597   &   70 56 30.764   & $0.66$ & 26.6 & $-4\pm$1.5 & $-0.5\pm$0.7 & $0.9\pm$0.8 \cr
Cl0841 & 4 & 19.4 & 0.2403 &   08 42 17.150   &   70 45 21.892   & ... & 8.8 & $-0.3\pm$4.2 & $1.7\pm$2.0 & $4.5\pm$1.9 \cr
Cl0841 & 10 & 20.0 & 0.2362 &   08 41 09.943   &   70 45 43.974   & ... & 10.4 & $9.7\pm$3.5 & $2.9\pm$1.6 & $27.6\pm$3.3 \cr
Cl0841 & 14 & 20.2 & 0.2423 &   08 41 43.495   &   70 46 24.265   & $0.78$ & 7 & $-2.7\pm$5.1 & $2.8\pm$2.4 & $-2.7\pm$2.6 \cr
Cl0841 & 15 & 19.1 & 0.2397 &   08 41 46.827   &   70 46 28.524   & $0.89$ & 15.4 & $0.8\pm$2.6 & $-0.2\pm$1.2 & $-2.5\pm$1.1 \cr
Cl0841 & 16 & 19.3 & 0.2403 &   08 41 47.226   &   70 46 31.627   & $0.91$ & 11 & $1.9\pm$3.9 & $-0.3\pm$1.7 & $-4.1\pm$1.5 \cr
Cl0841 & 20 & 18.3 & 0.2372 &   08 41 37.445   &   70 46 55.657   & $0.42$ & 29.8 & $0.2\pm$1.2 & $-0.0\pm$0.6 & $-1.5\pm$0.6 \cr
Cl0841 & 23 & 19.6 & 0.2396 &   08 41 14.286   &   70 46 59.340   & ... & 16.4 & $14.8\pm$2.3 & $4.3\pm$1.0 & $42.5\pm$2.5 \cr
Cl0841 & 24 & 20.0 & 0.2415 &   08 41 41.982   &   70 46 54.095   & $0.86$ & 5.2 & $4.8\pm$7.5 & $1.4\pm$3.5 & $-4.2\pm$3.1 \cr
Cl0841 & 25$^{a}$ & 17.9 & 0.2397 &   08 41 44.081   &   70 46 52.748   & $0.46$ & 29.8 & $1.8\pm$1.9 & $-1.4\pm$0.6 & $-2.7\pm$0.6 \cr
Cl0841 & 27 & 20.6 & 0.2384 &   08 41 40.979   &   70 47 15.598   & $0.63$ & 4 & $13.6\pm$13.4 & $3.7\pm$4.3 & $-4.2\pm$3.9 \cr
Cl0841 & 28 & 19.9 & 0.2367 &   08 41 04.120   &   70 47 15.572   & ... & 9.4 & $-5.2\pm$4.0 & $2.5\pm$1.8 & $0.0\pm$2.1 \cr
Cl0841 & 30 & 20.5 & 0.2396 &   08 41 54.499   &   70 47 31.308   & ... & 6.6 & $20.6\pm$6.1 & $-0.7\pm$2.8 & $36.0\pm$5.8 \cr
Cl0841 & 33 & 19.7 & 0.2401 &   08 41 25.184   &   70 47 45.895   & ... & 10.4 & $-9.1\pm$3.3 & $0.1\pm$1.7 & $-3.4\pm$1.7 \cr
Cl0841 & 34 & 19.2 & 0.2388 &   08 41 28.476   &   70 47 47.350   & ... & 14.2 & $-1\pm$2.8 & $0.5\pm$1.3 & $3.0\pm$1.3 \cr
Cl0841 & 37 & 18.7 & 0.2373 &   08 41 30.923   &   70 47 52.732   & $0.55$ & 21.6 & $-1.6\pm$1.7 & $0.1\pm$0.8 & $6.4\pm$0.9 \cr
Cl0841 & 38 & 18.1 & 0.2402 &   08 41 32.395   &   70 47 57.401   & $0.82$ & 34 & $11.7\pm$1.2 & $2.0\pm$0.5 & $10.6\pm$0.7 \cr
Cl0841 & 41 & 18.8 & 0.2415 &   08 40 48.377   &   70 48 26.352   & ... & 19.6 & $0.5\pm$2.0 & $0.7\pm$0.9 & $-2.1\pm$1.0 \cr
Cl0841 & 44 & 19.3 & 0.2396 &   08 41 50.230   &   70 48 43.243   & $0.63$ & 10.4 & $-2.3\pm$3.8 & $0.6\pm$1.8 & $-4.5\pm$1.8 \cr
Cl0841 & 46 & 19.8 & 0.2387 &   08 41 54.697   &   70 48 53.240   & $0.90$ & 7 & $-10.1\pm$5.8 & $0.6\pm$2.6 & $1.6\pm$2.8 \cr
Cl0841 & 50 & 19.9 & 0.2400 &   08 42 47.915   &   70 49 08.648   & ... & 5.8 & $-3.3\pm$5.9 & $0.4\pm$3.0 & $12.6\pm$3.6 \cr
\smallskip
Cl0841 & 51 & 19.5 & 0.2418 &   08 40 52.291   &   70 49 17.299   & ... & 9.8 & $19.3\pm$5.1 & $1.6\pm$2.3 & $118.3\pm$64.2 \cr
Cl0849 & 6 & 19.2 & 0.2346 &   08 49 15.608   &   37 28 14.408   & ... & 9 & ... & $-4.5\pm$2.2 & $1.4\pm$2.0 \cr
Cl0849 & 12 & 19.5 & 0.2295 &   08 49 07.537   &   37 29 00.082   & ... & 8.2 & $-0.8\pm$7.1 & $3.0\pm$3.0 & ... \cr
Cl0849 & 20 & 19.9 & 0.2382 &   08 49 07.650   &   37 30 25.679   & $0.90$ & 5.6 & $13.3\pm$9.5 & $6.4\pm$3.2 & ... \cr
Cl0849 & 21 & 18.7 & 0.2375 &   08 49 06.456   &   37 30 37.134   & ... & 13.2 & $1.6\pm$2.8 & $2.0\pm$1.3 & $6.1\pm$1.5 \cr
Cl0849 & 22 & 19.3 & 0.2347 &   08 49 06.888   &   37 30 47.131   & $0.35$ & 5.2 & $15.8\pm$7.6 & $-0.2\pm$3.5 & $40.1\pm$6.4 \cr
Cl0849 & 23 & 19.4 & 0.2315 &   08 49 09.719   &   37 30 59.368   & $0.83$ & 5 & $0.6\pm$7.2 & $3.4\pm$3.5 & $1.0\pm$4.5 \cr
Cl0849 & 24$^{a}$ & 17.9 & 0.2370 &   08 49 10.759   &   37 31 09.023   & $0.49$ & 23 & $0.5\pm$1.5 & $0.2\pm$0.8 & $-0.2\pm$0.7 \cr
\noalign{\smallskip}
\noalign{\hrule}
\end{tabular}
\end{center} 
\end{table*}
}

\setcounter{table}{3}
{\scriptsize
\begin{table*} 
\begin{center} 
\caption{continued} 
\vspace{0.1cm}
\begin{tabular}{lcccccccrrr} 
\hline\hline
\noalign{\smallskip}
Cluster&  Galaxy ID & $R_{\rm mag}$& Redshift & R.A.             & Dec          & B/T     & S/N  &\ewoii & \ewhd & \ewha \cr
       &            &              &          &\multispan{2}{\hfil (J2000)\hfil}&         &      &{\AA\hfil}   & {\AA\hfil}   &{\AA\hfil} \cr 
\noalign{\smallskip}
\hline
Cl0849 & 25 & 19.1 & 0.2310 &   08 49 10.506   &   37 31 14.898   & $0.47$ & 5.8 & $0.7\pm$6.4 & $1.0\pm$3.1 & $2.2\pm$4.4 \cr
Cl0849 & 26 & 18.3 & 0.2385 &   08 49 10.975   &   37 30 52.556   & $0.71$ & 18 & $-1.2\pm$2.2 & $-1.2\pm$1.1 & $-0\pm$1.0 \cr
Cl0849 & 27 & 18.6 & 0.2403 &   08 49 09.152   &   37 31 23.484   & ... & 16.4 & $0.3\pm$2.4 & $2.8\pm$1.1 & $-2.8\pm$1.2 \cr
Cl0849 & 29 & 18.5 & 0.2290 &   08 49 12.967   &   37 31 49.714   & ... & 16.6 & $-0.9\pm$2.3 & $0.9\pm$1.1 & ... \cr
Cl0849 & 34 & 19.0 & 0.2382 &   08 49 08.738   &   37 32 30.113   & $0.75$ & 10.4 & $-2.5\pm$4.0 & $2.1\pm$1.7 & $-1.1\pm$1.7 \cr
Cl0849 & 35 & 19.1 & 0.2391 &   08 49 11.554   &   37 32 37.036   & ... & 7.2 & $-1.6\pm$4.9 & $0.1\pm$2.4 & $-1.6\pm$3.1 \cr
Cl0849 & 37 & 19.7 & 0.2315 &   08 49 08.189   &   37 33 49.010   & ... & 5 & $0.6\pm$7.2 & $3.4\pm$3.5 & $1.0\pm$4.5 \cr
Cl0849 & 38 & 18.6 & 0.2335 &   08 49 00.132   &   37 33 53.186   & ... & 13.6 & $-0.1\pm$2.7 & $-1.4\pm$1.4 & $1.2\pm$1.5 \cr
Cl0849 & 39 & 18.4 & 0.2339 &   08 49 10.045   &   37 34 02.564   & ... & 11.6 & $0.2\pm$2.8 & $-1.2\pm$1.6 & $3.1\pm$1.9 \cr
Cl0849 & 40 & 19.5 & 0.2315 &   08 49 08.437   &   37 34 05.214   & ... & 6.2 & $3.6\pm$7.1 & $1.3\pm$3.0 & ... \cr
Cl0849 & 42 & 18.7 & 0.2318 &   08 49 09.584   &   37 34 18.426   & ... & 12.4 & $-3.7\pm$3.2 & $-1.5\pm$1.5 & $0.4\pm$1.5 \cr
Cl0849 & 45 & 19.5 & 0.2321 &   08 49 11.096   &   37 34 31.570   & ... & 6.4 & $3.6\pm$7.6 & $-3.0\pm$3.0 & $5.6\pm$3.1 \cr
Cl0849 & 46 & 19.7 & 0.2320 &   08 49 20.768   &   37 34 44.202   & ... & 5.2 & $1.8\pm$9.3 & $1.0\pm$3.4 & $-1.5\pm$4.0 \cr
Cl0849 & 50 & 19.0 & 0.2326 &   08 49 00.897   &   37 34 59.570   & ... & 10.2 & $-4.5\pm$4.6 & $-0.4\pm$1.8 & $-1.1\pm$1.9 \cr
Cl0849 & 52 & 19.9 & 0.2343 &   08 49 19.087   &   37 35 24.576   & ... & 6.2 & $14.6\pm$8.1 & $3.6\pm$2.8 & $23.9\pm$5.5 \cr
Cl0849 & 53 & 19.0 & 0.2346 &   08 49 04.545   &   37 35 44.873   & ... & 12 & $-4.5\pm$3.2 & $0.2\pm$1.5 & ... \cr
Cl0849 & 54 & 19.4 & 0.2337 &   08 49 13.623   &   37 36 00.119   & ... & 8.2 & $-9.2\pm$5.4 & $-2.0\pm$2.4 & $-5.4\pm$2.4 \cr
Cl0849 & 56 & 19.3 & 0.2356 &   08 49 09.393   &   37 36 12.438   & ... & 8.6 & $-2.5\pm$4.1 & $1.0\pm$2.1 & $1.1\pm$2.6 \cr
\smallskip
Cl0849 & -99 & 19.8 & 0.2309 &   08 49 09.635   &   37 35 42.871   & ... & 6 & $-4.2\pm$6.3 & $1.5\pm$3.0 & $-0.1\pm$4.4 \cr
Cl1309 & 48 & 19.7 & 0.2998 &   13 10 02.322   &   32 21 12.229   & ... & 11 & $11.3\pm$3.2 & $5.9\pm$1.4 & $26.9\pm$4.3 \cr
Cl1309 & 59 & 19.6 & 0.2961 &   13 10 06.826   &   32 21 22.957   & ... & 7 & $-3.9\pm$4.7 & $0.2\pm$2.5 & ... \cr
Cl1309 & 75 & 19.8 & 0.2907 &   13 10 04.200   &   32 21 34.204   & ... & 6.2 & $-2.9\pm$5.0 & $3.2\pm$2.6 & $-3.2\pm$4.9 \cr
Cl1309 & 81 & 19.8 & 0.2991 &   13 09 59.128   &   32 21 43.981   & ... & 4 & $3.4\pm$12.0 & $-0.2\pm$4.1 & ... \cr
Cl1309 & 83 & 19.2 & 0.2934 &   13 09 55.063   &   32 21 45.752   & $0.39$ & 9 & $-2.7\pm$4.0 & $-3.9\pm$2.0 & ... \cr
Cl1309 & 99 & 19.5 & 0.2996 &   13 09 53.225   &   32 21 56.326   & $0.80$ & 6.4 & $-3.3\pm$5.4 & $1.6\pm$2.6 & $0.7\pm$3.8 \cr
Cl1309 & 118 & 19.0 & 0.2930 &   13 10 11.341   &   32 22 01.092   & ... & 11 & $-5.9\pm$2.9 & $0.0\pm$1.6 & ... \cr
Cl1309 & 119 & 20.4 & 0.2977 &   13 09 53.093   &   32 22 10.801   & $0.26$ & 3.4 & ... & ... & $-12.1\pm$6.8 \cr
Cl1309 & 133 & 19.0 & 0.2927 &   13 09 51.577   &   32 22 14.070   & ... & 9.4 & $-5.3\pm$3.1 & $-2.3\pm$1.9 & $1.1\pm$2.7 \cr
Cl1309 & 142 & 19.8 & 0.2939 &   13 09 55.177   &   32 22 24.560   & $0.70$ & 5 & $0.6\pm$7.5 & $-0.8\pm$3.5 & ... \cr
Cl1309 & 151$^{a}$ & 18.4 & 0.2930 &   13 09 56.118   &   32 22 13.822   & $0.57$ & 19.2 & $-5.2\pm$1.7 & $-1.4\pm$0.9 & $3.2\pm$1.3 \cr
Cl1309 & 155 & 19.0 & 0.2934 &   13 09 58.502   &   32 22 28.668   & $0.73$ & 10.8 & $-3\pm$3.1 & ... & ... \cr
Cl1309 & 165 & 19.0 & 0.2945 &   13 10 13.066   &   32 22 49.667   & ... & 11.2 & $-1.2\pm$2.7 & $-0.0\pm$1.6 & ... \cr
Cl1309 & 174 & 19.4 & 0.2925 &   13 10 00.194   &   32 22 57.464   & ... & 12.8 & $6.3\pm$2.5 & $5.1\pm$1.2 & $13.6\pm$3.3 \cr
Cl1309 & 190 & 19.2 & 0.2913 &   13 09 56.349   &   32 23 08.066   & $0.47$ & 5.4 & $5.0\pm$6.1 & $1.2\pm$3.1 & $18.5\pm$5.4 \cr
Cl1309 & 191 & 19.9 & 0.2913 &   13 10 09.016   &   32 23 10.950   & ... & 5 & $3.6\pm$8.6 & $-0.7\pm$3.5 & $-5.5\pm$5.1 \cr
Cl1309 & 193 & 19.0 & 0.2927 &   13 09 57.689   &   32 23 10.525   & $0.62$ & 11 & $2.7\pm$3.3 & $-1.8\pm$1.6 & $3.0\pm$2.3 \cr
Cl1309 & 199 & 21.1 & 0.2933 &   13 09 42.056   &   32 23 29.558   & ... & 4.2 & $15.4\pm$7.7 & $-1.0\pm$4.3 & $97.5\pm$31.9 \cr
\smallskip
Cl1309 & 214 & 20.3 & 0.2942 &   13 09 45.026   &   32 23 31.866   & ... & 7.8 & $7.5\pm$4.1 & $2.8\pm$2.1 & ... \cr
Cl1444a & 74 & 19.5 & 0.2916 &   14 43 32.384   &   63 45 15.188   & ... & 6.8 & $-2.4\pm$5.7 & $-0.6\pm$2.5 & ... \cr
Cl1444a & 78 & 19.1 & 0.2899 &   14 43 54.807   &   63 45 13.885   & $0.42$ & 12 & $3.5\pm$2.7 & ... & ... \cr
Cl1444a & 80 & 18.2 & 0.2918 &   14 44 04.940   &   63 45 06.124   & ... & 17.6 & $-3\pm$1.9 & $-0.9\pm$1.0 & ... \cr
Cl1444a & 84 & 19.4 & 0.2921 &   14 43 46.399   &   63 45 16.726   & ... & 7.6 & $4.6\pm$5.2 & $0.5\pm$2.3 & ... \cr
Cl1444a & 89 & 19.4 & 0.2950 &   14 43 09.283   &   63 45 23.346   & ... & 8.4 & $3.5\pm$4.1 & $0.5\pm$2.1 & ... \cr
Cl1444a & 93 & 19.8 & 0.2920 &   14 44 15.062   &   63 45 23.468   & $0.00$ & 8.6 & $1.7\pm$3.8 & $4.4\pm$1.9 & $15.2\pm$8.4 \cr
Cl1444 & 100 & 18.7 & 0.2908 &   14 44 23.851   &   63 45 29.966   & ... & 14 & $0.1\pm$2.5 & $3.3\pm$1.2 & ... \cr
Cl1444 & 107 & 19.3 & 0.2949 &   14 43 50.343   &   63 45 41.130   & ... & 10 & $5.0\pm$3.3 & $3.1\pm$1.6 & $15.0\pm$4.1 \cr
Cl1444 & 110$^{a}$ & 18.0 & 0.2908 &   14 43 54.997   &   63 45 34.812   & $0.63$ & 23 & $0.1\pm$1.5 & $0.7\pm$0.8 & ... \cr
Cl1444 & 111 & 19.6 & 0.2902 &   14 44 01.033   &   63 45 41.612   & ... & 6.6 & $1.1\pm$4.8 & $1.3\pm$2.5 & ... \cr
Cl1444a & 115 & 19.0 & 0.2921 &   14 44 00.945   &   63 45 48.888   & ... & 13.2 & $-2.6\pm$2.6 & ... & ... \cr
Cl1444a & 121 & 19.2 & 0.2935 &   14 43 32.725   &   63 46 00.700   & ... & 11.2 & $9.9\pm$3.6 & $4.8\pm$1.4 & $6.2\pm$2.5 \cr
Cl1444 & 149 & 19.0 & 0.2951 &   14 43 16.978   &   63 46 37.916   & ... & 10.2 & $-3.3\pm$3.3 & $-1.3\pm$1.7 & ... \cr
Cl1444a & 170 & 19.1 & 0.2964 &   14 43 36.859   &   63 47 18.758   & ... & 12.2 & $5.2\pm$2.9 & $4.7\pm$1.3 & $13.2\pm$2.6 \cr
Cl1444a & 89 & 19.4 & 0.2950 &   14 43 09.283   &   63 45 23.346   & ... & 8.4 & $3.5\pm$4.1 & $0.5\pm$2.1 & ... \cr
\smallskip
Cl1444a & 149 & 19.0 & 0.2951 &   14 43 16.978   &   63 46 37.916   & ... & 10.2 & $-3.3\pm$3.3 & $-1.3\pm$1.7 & ... \cr
Cl1444a/b & 170 & 19.1 & 0.2964 &   14 43 36.859   &   63 47 18.758   & ... & 12.2 & $5.2\pm$2.9 & $4.7\pm$1.3 & $13.2\pm$2.6 \cr
Cl1444b & 24 & 18.8 & 0.2976 &   14 44 34.578   &   63 43 12.349   & ... & 11 & $7.8\pm$3.5 & $-0.7\pm$1.6 & ... \cr
Cl1444b & 28 & 19.9 & 0.2976 &   14 43 25.496   &   63 43 43.756   & ... & 4.6 & ... & $-0.7\pm$3.8 & ... \cr
Cl1444b & 59 & 18.7 & 0.3000 &   14 44 24.272   &   63 44 37.381   & ... & 11.6 & ... & $1.0\pm$1.5 & ... \cr
Cl1444b & 62 & 19.0 & 0.3014 &   14 44 06.299   &   63 44 57.185   & ... & 12.6 & $-1\pm$2.8 & $-0.9\pm$1.4 & ... \cr
Cl1444b & 70 & 18.8 & 0.3009 &   14 43 53.020   &   63 45 07.002   & $0.68$ & 16.8 & $-3.8\pm$1.9 & $0.4\pm$1.0 & ... \cr
\noalign{\smallskip}
\noalign{\hrule}
\end{tabular}
\end{center} 
\end{table*}
}

\setcounter{table}{3}
{\scriptsize
\begin{table*} 
\begin{center} 
\caption{continued} 
\vspace{0.1cm}
\begin{tabular}{lcccccccrrr} 
\hline\hline
\noalign{\smallskip}
Cluster&  Galaxy ID & $R_{\rm mag}$&  Redshift & R.A.             & Dec          & B/T     & S/N  &\ewoii & \ewhd & \ewha \cr
       &            &              &          &\multispan{2}{\hfil (J2000)\hfil}&         &      &{\AA\hfil}   & {\AA\hfil}   &{\AA\hfil} \cr 
\noalign{\smallskip}
\hline
Cl1444b & 71$^{a}$ & 18.4 & 0.2986 &   14 44 06.852   &   63 44 58.722   & ... & 13.4 & $1.3\pm$2.5 & $-0.7\pm$1.3 & ... \cr
Cl1444b & 75 & 18.2 & 0.3010 &   14 43 47.029   &   63 45 08.474   & ... & 22.8 & $-6.2\pm$1.4 & $-0.7\pm$0.8 & ... \cr
Cl1444b & 79 & 20.1 & 0.3017 &   14 43 53.265   &   63 45 16.067   & $0.56$ & 4.4 & $2.6\pm$6.3 & $0.8\pm$4.0 & ... \cr
Cl1444b & 85 & 19.3 & 0.2986 &   14 43 57.990   &   63 45 16.934   & $0.34$ & 7 & $3.3\pm$4.6 & $-2.7\pm$2.6 & ... \cr
Cl1444b & 86 & 20.2 & 0.2994 &   14 44 40.067   &   63 45 10.246   & ... & 3.6 & $4.4\pm$11.7 & $1.3\pm$4.8 & ... \cr
Cl1444b & 113 & 20.4 & 0.3023 &   14 44 09.141   &   63 45 46.721   & $1.00$ & 3.6 & $-11.1\pm$7.8 & $3.1\pm$4.9 & ... \cr
Cl1444b & 122 & 19.7 & 0.3045 &   14 43 40.895   &   63 46 02.305   & ... & 5.8 & $13.3\pm$6.1 & $-3.1\pm$3.2 & $7.5\pm$6.1 \cr
Cl1444b & 138 & 19.2 & 0.3017 &   14 43 57.286   &   63 46 18.332   & $0.77$ & 12.2 & $-5.7\pm$2.6 & $1.0\pm$1.5 & $-0.6\pm$2.1 \cr
Cl1444b & 157 & 18.8 & 0.3001 &   14 44 02.710   &   63 46 56.017   & $0.82$ & 14.6 & $-2.4\pm$2.5 & $0.2\pm$1.2 & ... \cr
\smallskip
Cl1444b & 166 & 19.9 & 0.3032 &   14 43 38.507   &   63 47 15.475   & ... & 6 & $9.1\pm$7.0 & $10.2\pm$2.7 & $22.6\pm$6.0 \cr
Cl1633 & 1 & 19.8 & 0.2373 &   16 33 28.890   &   57 12 10.577   & ... & 11.6 & $-0\pm$4.1 & $4.5\pm$2.0 & ... \cr
Cl1633 & 9 & 20.9 & 0.2412 &   16 33 29.194   &   57 13 37.340   & ... & 3.4 & $-7.6\pm$9.3 & $1.5\pm$5.3 & ... \cr
Cl1633 & 13 & 19.9 & 0.2450 &   16 33 16.311   &   57 14 04.877   & ... & 12.6 & $44.2\pm$3.7 & $3.4\pm$1.3 & $50.4\pm$5.9 \cr
Cl1633 & 14 & 20.1 & 0.2388 &   16 33 35.339   &   57 14 04.218   & $0.82$ & 6.6 & $-4.8\pm$5.1 & $-1.6\pm$2.7 & $-6.8\pm$3.5 \cr
Cl1633 & 15 & 19.8 & 0.2417 &   16 33 42.836   &   57 14 02.378   & $0.62$ & 5.4 & $3.2\pm$9.5 & $-3.0\pm$5.0 & ... \cr
Cl1633 & 17 & 20.2 & 0.2402 &   16 33 40.276   &   57 14 11.522   & $0.77$ & 5 & $4.9\pm$12.7 & $-2.6\pm$5.3 & ... \cr
Cl1633 & 22 & 19.1 & 0.2388 &   16 33 41.726   &   57 14 20.436   & $0.75$ & 10.2 & $-11.5\pm$4.7 & $-0.2\pm$2.5 & ... \cr
Cl1633 & 23 & 20.6 & 0.2422 &   16 33 43.374   &   57 14 27.013   & $0.77$ & 4.6 & $0.1\pm$6.8 & $1.0\pm$3.9 & $0.4\pm$7.2 \cr
Cl1633 & 24 & 21.0 & 0.2417 &   16 33 14.114   &   57 14 33.799   & ... & 4.2 & $46.3\pm$12.2 & $-4.2\pm$4.7 & $45.7\pm$18.5 \cr
Cl1633 & 25$^{a}$ & 18.4 & 0.2402 &   16 33 42.129   &   57 14 12.649   & $0.56$ & 18 & $-2.3\pm$1.7 & $-1.0\pm$1.0 & ... \cr
Cl1633 & 27 & 19.8 & 0.2396 &   16 33 31.879   &   57 14 31.517   & ... & 7.6 & $-2.2\pm$4.9 & $0.5\pm$2.4 & $0.0\pm$3.1 \cr
Cl1633 & 28 & 20.3 & 0.2371 &   16 33 30.890   &   57 14 31.585   & ... & 9 & $16.9\pm$6.5 & $-0.4\pm$2.9 & ... \cr
Cl1633 & 31 & 19.4 & 0.2425 &   16 33 41.155   &   57 14 29.281   & $0.71$ & 9.8 & $-5.3\pm$3.4 & $-2.8\pm$1.9 & ... \cr
Cl1633 & 34 & 19.1 & 0.2407 &   16 33 05.867   &   57 14 52.379   & ... & 13 & $-0.4\pm$2.8 & $0.0\pm$1.4 & ... \cr
Cl1633 & 35 & 19.2 & 0.2366 &   16 33 50.398   &   57 14 46.705   & ... & 12.6 & $1.0\pm$2.8 & $-0.0\pm$1.5 & ... \cr
Cl1633 & 43 & 20.7 & 0.2404 &   16 33 39.426   &   57 15 14.144   & $0.43$ & 4.2 & $-1.5\pm$7.7 & $-0.0\pm$4.5 & ... \cr
Cl1633 & 46 & 18.8 & 0.2384 &   16 33 24.291   &   57 15 11.686   & ... & 18.4 & $5.3\pm$1.9 & $2.3\pm$1.0 & ... \cr
\smallskip
Cl1633 & 47 & 19.2 & 0.2394 &   16 33 56.034   &   57 15 23.566   & ... & 12.6 & $-3.9\pm$2.7 & $0.2\pm$1.5 & ... \cr
Cl1701 & 42 & 20.1 & 0.2464 &   17 01 59.839   &   64 19 21.036   & ... & 5.6 & $5.2\pm$6.2 & $3.8\pm$2.7 & $13.8\pm$4.9 \cr
Cl1701 & 86 & 19.5 & 0.2480 &   17 01 25.239   &   64 20 14.593   & ... & 6.8 & $4.0\pm$5.5 & $3.8\pm$2.3 & $4.5\pm$3.1 \cr
Cl1701 & 102 & 19.4 & 0.2422 &   17 01 27.279   &   64 20 47.692   & ... & 13.6 & $30.4\pm$3.1 & $5.3\pm$1.2 & $53.2\pm$3.7 \cr
Cl1701 & 116 & 19.7 & 0.2460 &   17 02 17.212   &   64 20 59.114   & ... & 6.6 & $3.3\pm$7.0 & $0.7\pm$2.7 & $-4.1\pm$2.5 \cr
Cl1701 & 118 & 19.8 & 0.2475 &   17 01 49.471   &   64 21 08.316   & $0.65$ & 6.2 & $5.0\pm$7.5 & ... & $-6.7\pm$2.5 \cr
Cl1701 & 120 & 19.3 & 0.2428 &   17 01 56.814   &   64 21 02.549   & $1.00$ & 8.6 & $-0.9\pm$4.1 & $1.3\pm$2.0 & $-2.8\pm$2.1 \cr
Cl1701 & 122 & 19.1 & 0.2451 &   17 01 36.610   &   64 21 07.960   & ... & 11.8 & $5.7\pm$3.5 & $0.9\pm$1.5 & $7.8\pm$1.7 \cr
Cl1701 & 123$^{a}$ & 18.4 & 0.2457 &   17 01 47.754   &   64 21 00.133   & $0.86$ & 18.6 & $-2.2\pm$1.9 & $-1.4\pm$1.0 & $-4.8\pm$1.0 \cr
Cl1701 & 124 & 18.4 & 0.2408 &   17 01 46.794   &   64 20 57.304   & $0.48$ & 16.6 & $0.7\pm$2.3 & $-1.0\pm$1.1 & $-0.3\pm$0.9 \cr
Cl1701 & 127 & 19.2 & 0.2411 &   17 02 06.423   &   64 21 12.629   & ... & 8.4 & $-0.9\pm$4.3 & $3.5\pm$2.0 & ... \cr
Cl1701 & 141 & 19.0 & 0.2429 &   17 01 51.826   &   64 21 32.677   & $0.38$ & 11.6 & $43.1\pm$4.8 & $-0.5\pm$1.5 & $19.2\pm$1.6 \cr
\smallskip
Cl1701 & 149 & 19.3 & 0.2458 &   17 01 42.825   &   64 21 49.460   & $0.28$ & 11.4 & $3.8\pm$3.2 & $4.6\pm$1.4 & $23.4\pm$2.5 \cr
Cl1702 & 14 & 19.2 & 0.2193 &   17 02 43.737   &   64 18 05.504   & ... & 9.8 & $3.4\pm$4.6 & $-0.8\pm$1.8 & $10.3\pm$2.5 \cr
Cl1702 & 22 & 20.3 & 0.2233 &   17 02 23.804   &   64 18 36.155   & ... & 3.8 & $-12.4\pm$10.4 & $3.4\pm$4.5 & $1.6\pm$6.6 \cr
Cl1702 & 28 & 18.6 & 0.2250 &   17 01 56.744   &   64 18 40.219   & ... & 17.2 & $3.6\pm$2.1 & $0.3\pm$1.0 & $8.0\pm$1.1 \cr
Cl1702 & 35 & 18.2 & 0.2225 &   17 01 40.986   &   64 18 56.372   & ... & 21 & $-1.6\pm$1.9 & $0.4\pm$0.8 & $-0.5\pm$0.8 \cr
Cl1702 & 41 & 19.4 & 0.2233 &   17 01 40.272   &   64 19 19.855   & ... & 7.6 & $1.1\pm$5.0 & $3.0\pm$2.1 & $-0.3\pm$2.7 \cr
Cl1702 & 60 & 19.3 & 0.2233 &   17 02 13.978   &   64 19 45.426   & $0.95$ & 8 & $-10.2\pm$4.5 & $-1.8\pm$2.2 & ... \cr
Cl1702 & 66 & 19.0 & 0.2225 &   17 02 08.958   &   64 19 51.852   & ... & 9 & ... & $-0.0\pm$1.9 & ... \cr
Cl1702 & 72 & 18.7 & 0.2212 &   17 02 14.817   &   64 19 46.412   & $0.76$ & 10.4 & $2.7\pm$3.5 & $1.2\pm$1.7 & $4.7\pm$1.9 \cr
Cl1702 & 81$^{a}$ & 17.6 & 0.2230 &   17 02 13.923   &   64 19 53.555   & $0.58$ & 38.4 & $-1.4\pm$1.0 & $-0.4\pm$0.5 & $-1.4\pm$0.5 \cr
Cl1702 & 87 & 19.3 & 0.2257 &   17 02 09.499   &   64 20 14.896   & ... & 9.2 & $-1.1\pm$4.6 & $1.3\pm$1.9 & $0.9\pm$1.9 \cr
Cl1702 & 90 & 18.0 & 0.2219 &   17 02 06.892   &   64 20 07.451   & ... & 22.4 & $-0.3\pm$1.8 & $-0.2\pm$0.8 & $-1.7\pm$0.8 \cr
Cl1702 & 129 & 18.5 & 0.2238 &   17 02 31.282   &   64 21 09.526   & ... & 9.2 & $-3.7\pm$3.9 & $3.3\pm$1.8 & ... \cr
Cl1702 & 130 & 18.6 & 0.2236 &   17 01 36.647   &   64 21 15.678   & ... & 15.2 & $0.9\pm$2.6 & $-1.8\pm$1.2 & $-3.8\pm$1.3 \cr
Cl1702 & 138 & 19.7 & 0.2243 &   17 02 04.065   &   64 21 37.210   & ... & 7.4 & $-4.1\pm$5.1 & $-1.2\pm$2.5 & $-4.1\pm$2.6 \cr
Cl1702 & 143 & 18.4 & 0.2224 &   17 02 01.652   &   64 21 38.531   & ... & 18.8 & $-1.3\pm$2.1 & $0.7\pm$0.9 & ... \cr
\noalign{\smallskip}
\noalign{\hrule}
\end{tabular}
\end{center} 
\begin{flushleft}
$^{a}$Central, giant elliptical galaxy.\\
\end{flushleft}
\end{table*}
}
\end{document}